\documentclass[fleqn,usenatbib]{mnras}

\usepackage[T1]{fontenc}
\usepackage{ae,aecompl}
\DeclareRobustCommand{\VAN}[3]{#2}
\let\VANthebibliography\thebibliography
\def\thebibliography{\DeclareRobustCommand{\VAN}[3]{##3}\VANthebibliography}

\usepackage{graphicx}	
\usepackage{amsmath}	
\usepackage{amssymb}	
\usepackage{gensymb}
\usepackage{siunitx}
\usepackage{array}
\usepackage{float}
\usepackage{amsfonts}
\usepackage{physics}
\usepackage{siunitx}
\usepackage{tikz,newtxmath}
\usepackage{mathtools}
\usepackage{mhchem}
\usepackage{color}
\usepackage{xcolor}
\usepackage{listings}
\usepackage{booktabs}
\usepackage{makecell}
\usepackage{subcaption}
\usepackage{tabularx}
\usepackage[T1]{fontenc}
\usepackage{longtable}
\usepackage{afterpage}
\usepackage{tablefootnote}
\usepackage[flushleft]{threeparttable}
\usepackage{soul}

\setstcolor{red}

\DeclareSIUnit \parsec {pc}
\DeclareSIUnit \year {yr}
\DeclareSIUnit \erg {erg}
\DeclareSIUnit \gauss {g}
\newcommand{\ergs}{\si{\erg}}
\newcommand{\ergspersecond}{\si{\erg\per\second}}
\newcommand{\pc}{\si{\parsec}}
\newcommand{\kpc}{\si{\kilo\parsec}}
\newcommand{\eV}{\si{\electronvolt}}
\newcommand{\GeV}{\si{\giga\electronvolt}}
\newcommand{\TeV}{\si{\tera\electronvolt}}

\newcommand{\centimeterminusthree}{\si{\per\centi\meter\cubed}}

\newcommand{\keV}{\si{\kilo\electronvolt}}

\newcommand{\kmpersec}{\si{\kilo\meter\per\second}}
\newcommand{\kiloyear}{\si{\kilo\year}}

\newcommand{\proton}{\ce{p}}
\newcommand{\electron}{\ce{e^-}}

\newcommand{\pionplus}{\ce{\pi^+}}
\newcommand{\pionminus}{\ce{\pi^-}}
\newcommand{\pionneutral}{\ce{\pi^0}}

\newcommand{\HESS}{H.E.S.S.}
\newcommand{\fermi}{\mbox{\emph{Fermi}}}
\newcommand{\halpha}{H$\alpha$ }

\newcommand{\HESSsource}{\mbox{HESS\,J1825-137} }

\title[GeV $\gamma$-rays adjacent to \HESSsource]{Explaining the extended $\GeV$ gamma-ray emission adjacent to \HESSsource}

\author[T. Collins et al.]{
T. Collins,$^{1}$\thanks{E-mail: tiffany.collins@adelaide.edu.au}, G. Rowell$^{1}$, A.M.W. Mitchell$^{2}$, F. Voisin$^{1}$, Y. Fukui$^{3}$, H. Sano$^{3}$, \newauthor R. Alsulami$^{1,4}$ and S. Einecke${^1}$
 \\
$^{1}$School of Physical Sciences, University of Adelaide, Adelaide 5005, Australia\\
$^{2}$Department of Physics, ETH Zurich, CH-8093 Zurich, Switzerland\\
$^{3}$Department of Physics, University of Nagoya, Furo-cho, Chikusa-ku, Nagoya, 464-8601, Japan \\
$^{4}$Astronomy Dept, Faculty of Science, King Abdulaziz University, Jeddah, Saudi Arabia\\
}

\date{Accepted XXX. Received YYY; in original form ZZZ}

\pubyear{2020}

\usepackage{newtxtext,newtxmath}

\begin{document}
\label{firstpage}
\pagerange{\pageref{firstpage}--\pageref{lastpage}}
\maketitle

\begin{abstract}
\HESSsource is one of the most powerful and luminous TeV gamma-ray pulsar wind nebulae (PWN). To the south of HESS\,J1825-137, \fermi-LAT observation revealed a new region of $\GeV$ gamma-ray emission with three apparent peaks (termed here, GeV-ABC). This study presents interstellar medium (ISM) data and spectral energy distribution (SED) modelling towards the $\GeV$ emission to understand the underlying particle acceleration. We considered several particle accelerator scenarios - the PWN associated with \HESSsource, the progenitor SNR also associated with HESS\,J1825-137, plus the gamma-ray binary system LS\,5039. It was found that the progenitor SNR of HESS\,J1825-137 has insufficient energetics to account for all $\GeV$ emission. GeV-ABC may be a reflection of an earlier epoch in the history of the PWN associated with HESS\,1825-137, assuming fast diffusion perhaps including advection. LS\,5039 cannot meet the required energetics to be the source of particle acceleration. A combination of \HESSsource and LS\,5039 could be plausible sources.
\end{abstract}




\section{Introduction}

\begin{figure}
    \centering
    \includegraphics[width=\columnwidth]{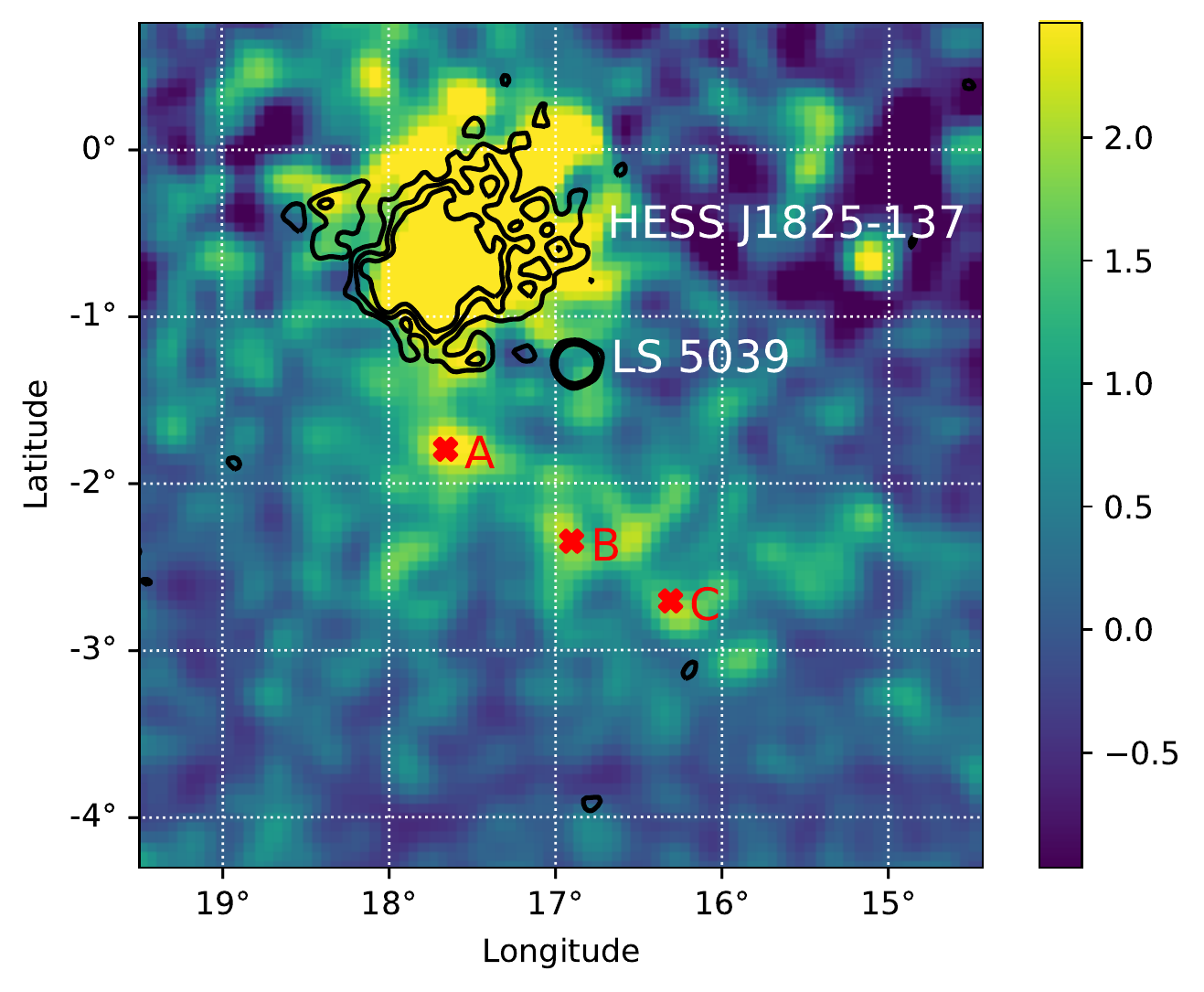}
    \caption{\fermi-LAT count map above $10~\GeV$ \citep{Araya:2019dez} towards \HESSsource overlaid by black HESS significance contours at $1\sigma$, $2\sigma$ and $3\sigma$ \citep{2018A&A...612A...1H}.}
    \label{fig:energy_HESS}
\end{figure}

HESS\,J1825-137 is a luminous pulsar wind nebula (PWN) powered  by the pulsar PSR J1826-1334 with spin down power of $\dot{E}=2.8\times 10^{36}~\ergspersecond$ and characteristic age of $21.4~\kiloyear$ \citep{2006A&A...460..365A,2005AJ....129.1993M}. To the south of HESS\,J1825-137 a new region of GeV gamma-ray emission was revealed by \protect\cite{Araya:2019dez} using \fermi-LAT data (see Fig.\,1). \protect\cite{Araya:2019dez} also suggested that this new region of GeV emission may be either an extension of HESS\,J1825-137 or unrelated to the system. If related to HESS\,J1825-137, the gamma-rays may be resultant from high energy particles from the PWN (in the form of electrons and positrons) or from the progenitor supernova remnant (SNR) linked to HESS\,J1825-137. If unrelated to HESS\,J1825-137 another source of high energy particles must exist towards this region. \cite{Araya:2019dez} conducted spectral analysis in range $10-250~\GeV$ and fitted the spectra observed from this new region to a power-law ($\dv{N}{E}\propto E^{-\Gamma}$) with index $\Gamma = 1.92\pm 0.07_\text{stat}\pm 0.05_\text{sys}$. Fig.\,3 from \cite{Araya:2019dez} shows a TS map towards this region with three distinct peaks. We label the three $\GeV$ features GeV-A, GeV-B and GeV-C and are located at positions RA: $18^\text{h}29^\text{m}36.0^\text{s}$, Dec: $-14\degree23'41.6''$, RA: $18^\text{h}30^\text{m}10.6^\text{s}$, Dec: $-15\degree19'03.4''$ and RA: $18^\text{h}30^\text{m}21.4^\text{s}$, Dec: $-16\degree00'40.3''$ respectively. Hereafter, the extended region of GeV emission will be referred to as GeV-ABC for simplicity. Fig.\,\ref{fig:energy_HESS} shows the locations of GeV-ABC relative to HESS\,J1825-137.
\par 
A $\TeV$ halo may be associated with \HESSsource \citep{2020MNRAS.494.2618L}. \TeV~halos occur when electrons and positrons escape the PWN through diffusion and interact with the ambient interstellar medium producing surrounding $\TeV$ emission forming a `halo'. The equivalent HAWC observatory source, eHWC\,J1825-134, has detected an extension of $\ang{0.36}$ above $56~\TeV$ \cite{2019arXiv190908609H}. The extension around \HESSsource can be seen to decrease with energy as shown by \fermi-LAT data \cite{2019ICRC...36..595P}. It is possible that GeV-ABC may be an extension of the $\TeV$ emission around \HESSsource
\par 
A H$\alpha$ rim like structure has been noted $120~\pc$ to the south east of the pulsar from PSR\,J1826-1334 \citep{2016MNRAS.458.2813V}. This structure may be associated with the progenitor supernova remnant (SNR) that is linked to PSR\,J1826-1334. The H$\alpha$ region overlays the \fermi-LAT GeV emission.
\par 
Another potential accelerator also towards GeV-ABC is the gamma-ray binary system LS\,5039, comprising a compact object and a massive O-type star 
\par 
This study presents gas analysis (CO, HI and H$\alpha$) towards the new region of \fermi-LAT GeV emission. To identify the origin of the GeV emission, spectral energy distribution (SED) modelling of the gamma-ray emission is performed assuming hadronic or leptonic particle populations accelerated by continuous or impulsive particle injectors.

\section{Gas Morphology towards GeV-ABC} \label{sec:gas}

PSR\,J1826-1334 and LS-5039, located within the vicinity towards GeV-ABC, are possible particle accelerators to produce the $GeV$ emission as seen by \fermi-LAT. PSR\,J1826-1334 has measured dispersion distance of $3.9\pm0.4~\kpc$ \citep{1993ApJ...411..674T} while the binary system LS-5039 distance is estimated to be $2.54\pm 0.04~\kpc$ \citep{2005MNRAS.364..899C}. For these two reasons ISM data in the velocity range of $15-30~\kmpersec$ and $40-60~\kmpersec$ corresponding to distances $1.6-2.8~\kpc$ and $3.5-4.5~\kpc$ respectively will be examined \citep{1993A&A...275...67B}.

\subsection{CO data} \label{sec:CO_data}

\begin{figure}
	\begin{subfigure}[b]{\columnwidth}
        \includegraphics[width=\textwidth]{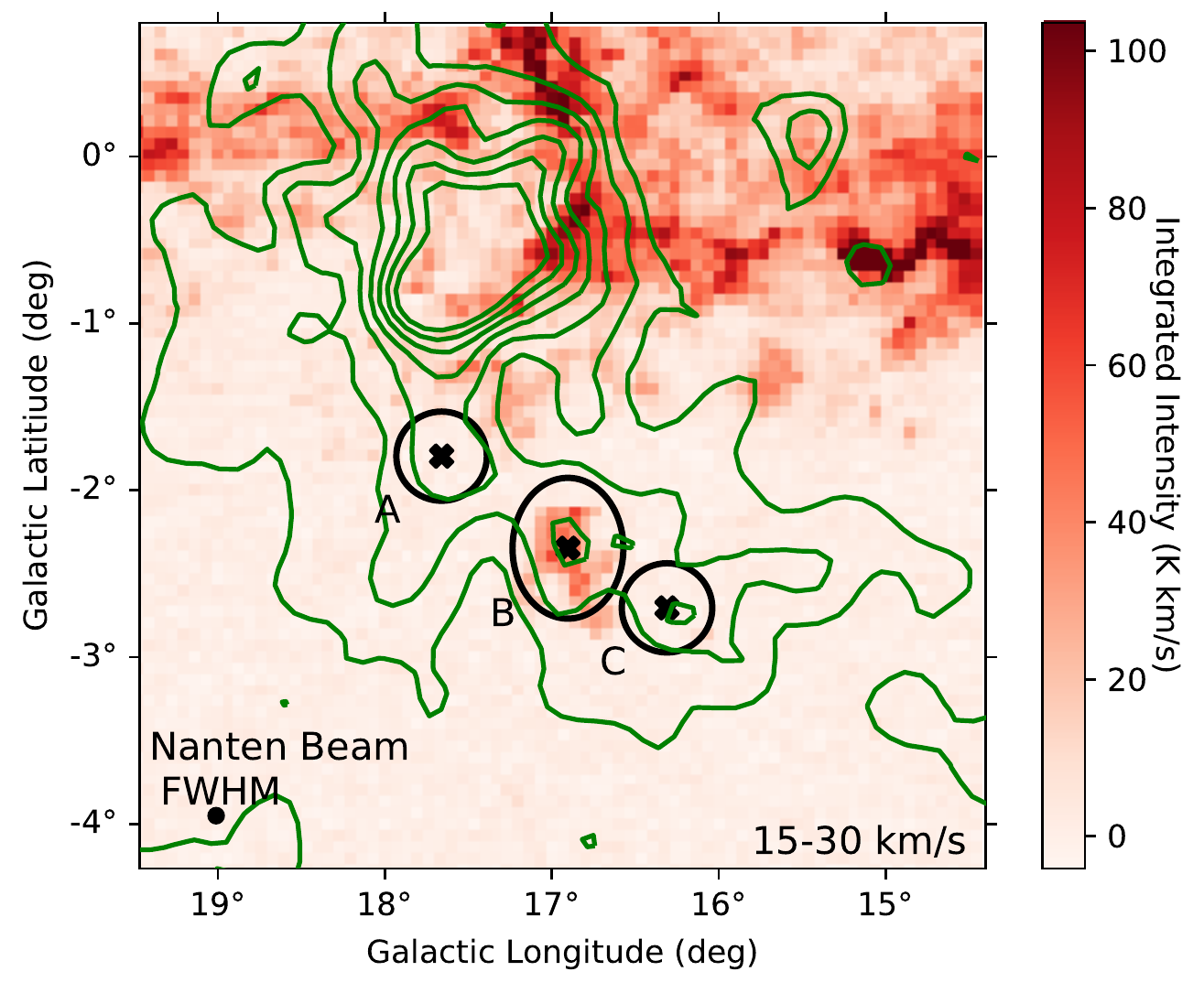}
	\end{subfigure}
	 
	\begin{subfigure}[b]{\columnwidth}
	    \includegraphics[width=\textwidth]{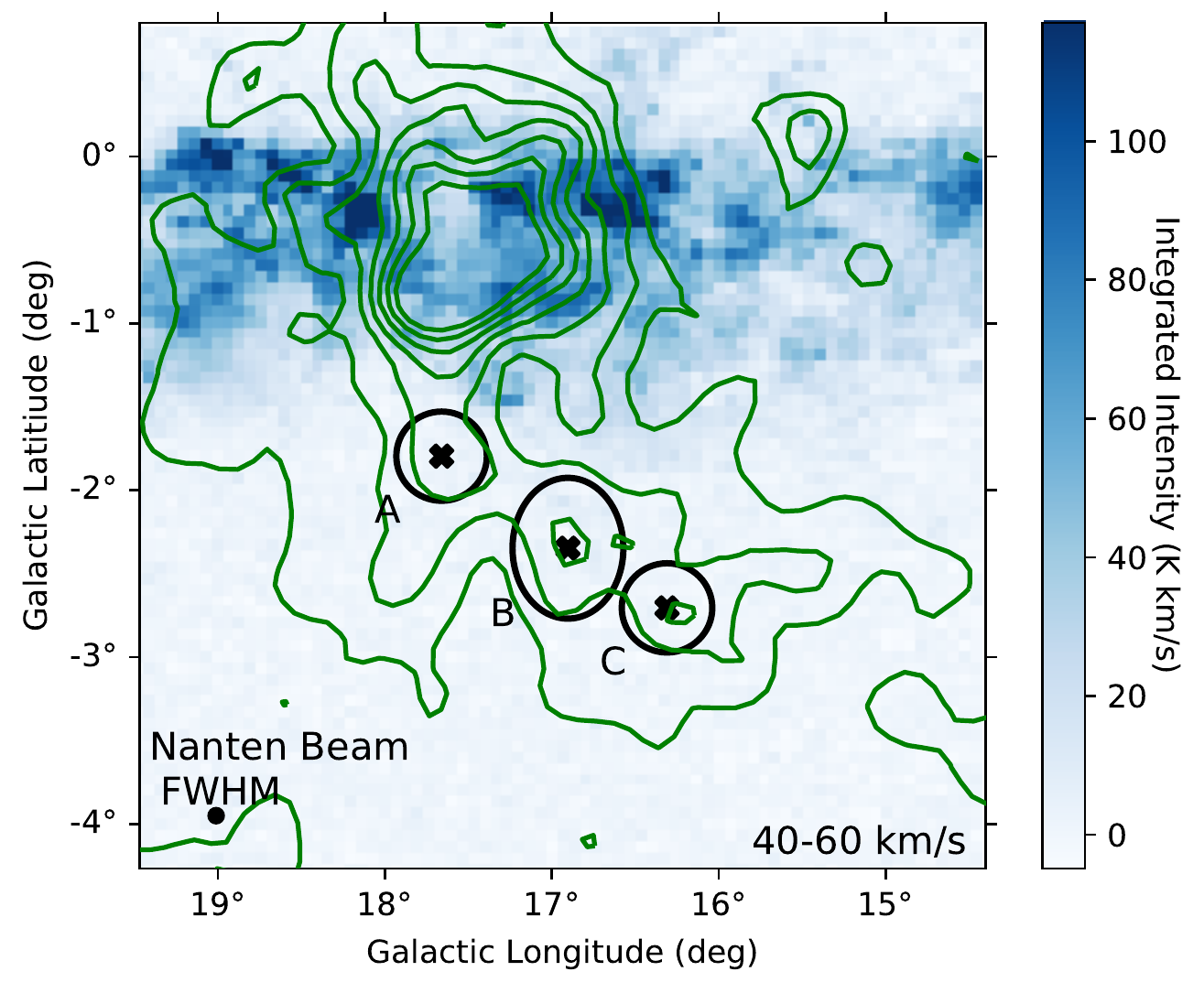}
	\end{subfigure}
    \caption{Nanten 12CO(1-0) integrated intensity in velocity ranges $15-30~\kmpersec$ (\emph{top}) and $40-60~\kmpersec$ (\emph{bottom}) . Green contours represents GeV emission as seen by \fermi-LAT at $1\sigma$ to $7\sigma$. The new regions of GeV emission, GeV-A, GeV-B and GeV-C, are shown by black markers. The Nanten beam size, shown in bottom left, is $2.6'$ \citep{2004ASPC..317...59M}.}
    \label{fig:CO_data}
\end{figure}

Using the Nanten 12CO(1-0) survey data, the molecular hydrogen column density will be traced using conversion factor $N_{H_2}=X_\text{CO}W_{12CO}$. The $X_\text{CO}$ factor is assumed to be constant $\approx 1.5\times 10^{20}~\si{\per\centi\meter\squared\per\kelvin\per\kilo\meter\second}$ \citep{2004A&A...422L..47S}, over the galactic plane but may vary with galactocentric radius.
\par 
The top panel in Fig.\,\ref{fig:CO_data} shows the 12CO(1-0) integrated intensity between $15-30~\kmpersec$. Regions of clumpy gas are noted to the north east of \HESSsource as noted by \cite{2016MNRAS.458.2813V}. Towards GeV-B, a region of denser gas are noticed which does not appear in the $40-60~\kmpersec$ range as shown in the bottom panel of Fig.\,\ref{fig:CO_data}. In both velocity ranges the region towards GeV-ABC has relatively little gas compared to the galactic plane. The distance to these clouds is determined from the galactic rotation curve. Individual gas motion may give a false interpretation of the velocity range. In \citep{1993A&A...275...67B} it was noticed that residuals of the modelled vs observed galactic rotation curve can be as great as $40~\kmpersec$ with the average being around $12.8~\kmpersec$.
\par 
The mass of a cloud with average column density $N_{H_2}$ and cloud area $A$ can be calculated by:

\begin{align}
    M_H&=2.8 N_{H_2} A \frac{m_p}{m_\odot}~{M_\odot}   
\end{align}

where $M_H = 2.8M_{H_2}$ includes a $20\%$ He component. The cloud areas used can be seen in Fig. \ref{fig:CO_data} The number density can then be obtained through:

\begin{align}
    n_H&= \frac{M_H}{4/3\pi R^3m_p}~\centimeterminusthree
\end{align}

where R represents the radius of cloud area considered. The results of these calculations are given in Table\,\ref{tab:CO_density} with cloud areas shown in Fig.\,\ref{fig:CO_data}. The size and shape of object B is chosen to contain dense gas seen in the $15-30~\kmpersec$ velocity ranges, while object A and C were chosen to be the same size but independent of object B.  The clumps seen towards GeV-B in the top panel of Fig.\,\ref{fig:CO_data} is an order of magnitude denser compared to the ISM towards GeV-A and GeV-C.

\subsection{HI data}

\begin{figure}
	\begin{subfigure}[b]{\columnwidth}
        \includegraphics[width=\textwidth]{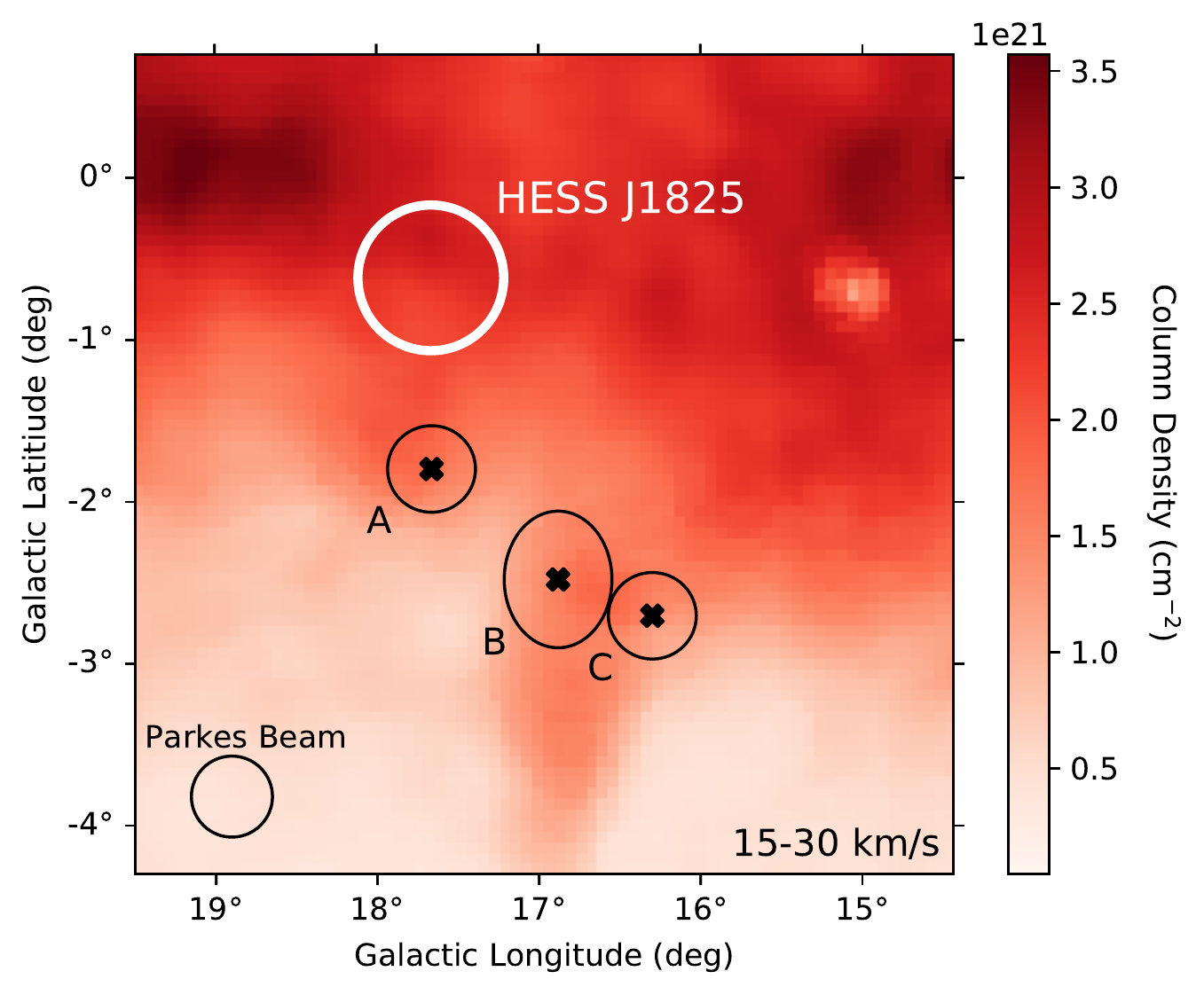}
	\end{subfigure}
	 
	\begin{subfigure}[b]{\columnwidth}
	    \includegraphics[width=\textwidth]{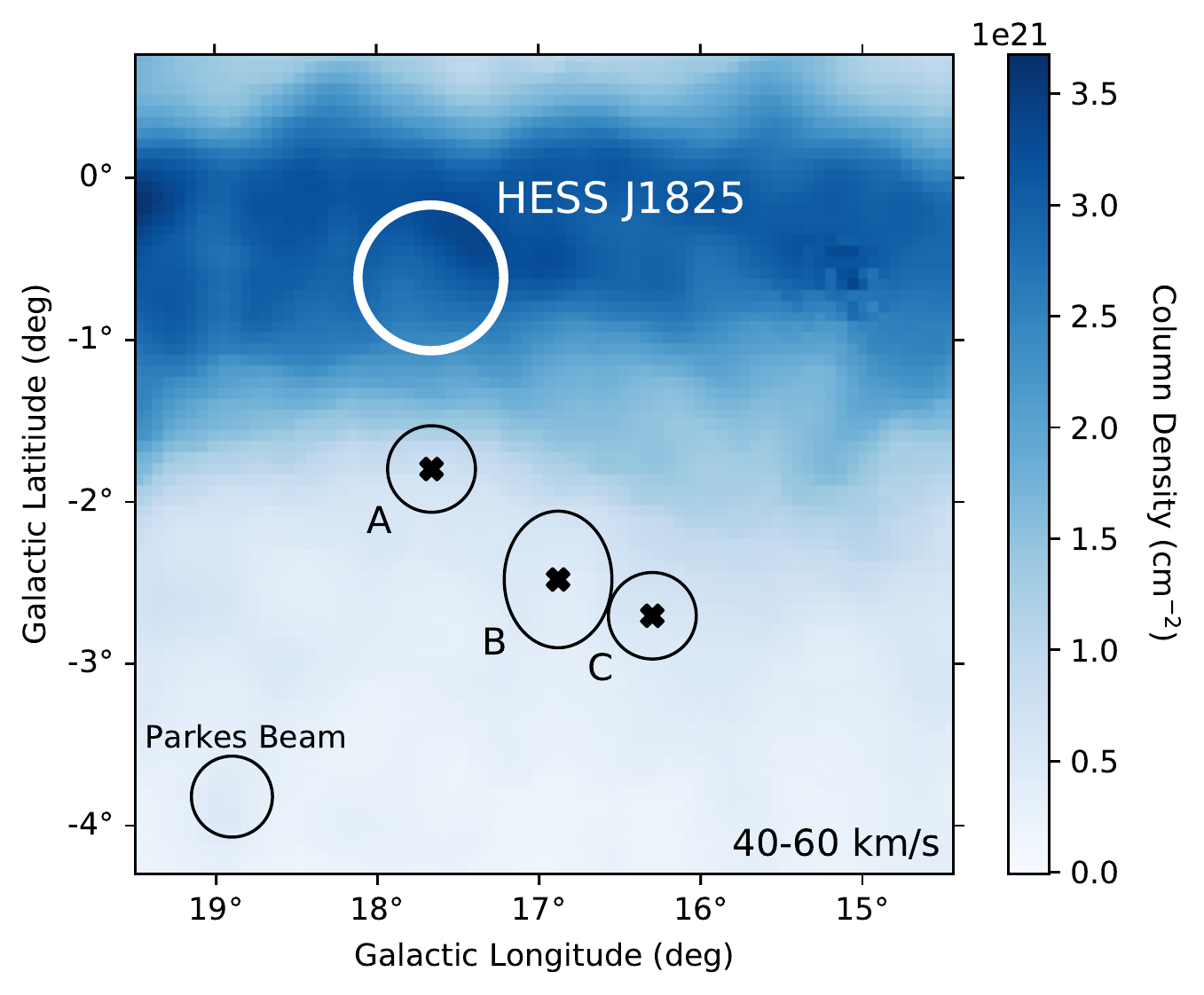}
	\end{subfigure}
    \caption{Parkes HI integrated column density (from GASS) in velocity ranges $15-30~\kmpersec$ (\emph{top}) and $40-60~\kmpersec$ (\emph{bottom}) \citep{2009ApJS..181..398M}. The regions that were analysed here can be seen in white (\HESSsource) and black (new GeV regions). The Parkes beam size, shown in bottom left, is $15'$ See Table\,\ref{tab:HI_density} for results.}
    \label{fig:HI_data}
\end{figure}

The Galactic All Sky Survey of atomic Hydrogen (HI) data set will be used to trace atomic hydrogen towards new region of GeV gamma-ray activity \citep{2009ApJS..181..398M}.
\par 
The integrated column density in the velocity of ranges of interest can be seen in Fig.\,\ref{fig:HI_data}. In the $40-60~\kmpersec$ range towards the region around GeV-ABC, the HI column density is relatively low compared to the galactic plane. The area towards the new GeV emission has slightly greater HI density in the $15-30~\kmpersec$ velocity range compared to the $40-60~\kmpersec$ velocity range.
\par 
The calculated HI parameters for different regions towards \HESSsource and the new emission of GeV gamma-rays can be seen in Table\,\ref{tab:HI_density}. Atomic hydrogen, compared to molecular hydrogen, is less abundant. The total ISM parameters are shown in Table\,\ref{tab:total_density}. The contribution of atomic hydrogen compared to molecular hydrogen is minimal (approximately $10\%$) to the total density of hydrogen gas.

\begin{table}
    \caption{Total ISM densities for \HESSsource and new GeV emission regions GeV-A, GeV-B and GeV-C.}
	\centering
	\begin{tabular}{cccc}
		\hline
		$15-30~\kmpersec$ & Object & $M_H$ ($M_\odot$) & $n_H$ ($\centimeterminusthree$) \\ 
	    \hline
	    & \HESSsource & $1.18\times 10^5$ & $40.1$ \\
	    & GeV-A & $4.56\times 10^3$ & $7.2$ \\
	    & GeV-B & $1.38\times 10^5$ & $79.8$ \\
	    & GeV-C & $2.30\times 10^3$ & $3.6$ \\
	    \hline
	    $40-60~\kmpersec$ & Object & $M_H$ ($M_\odot$) & $n_H$ ($\centimeterminusthree$) \\ 
	    \hline
	    & \HESSsource & $5.22\times 10^5$ & $178$ \\
	    & GeV-A & $8.93\times 10^3$ & $14$ \\
	    & GeV-B & $1.26\times 10^4$ & $7.3$ \\
	    & GeV-C & $3.11\times 10^2$ & $0.5$ \\
	    \hline
	\end{tabular}
	\label{tab:total_density}
\end{table}

\subsection{H$\alpha$ data}

\begin{figure}
    \centering
    \includegraphics[width=\columnwidth]{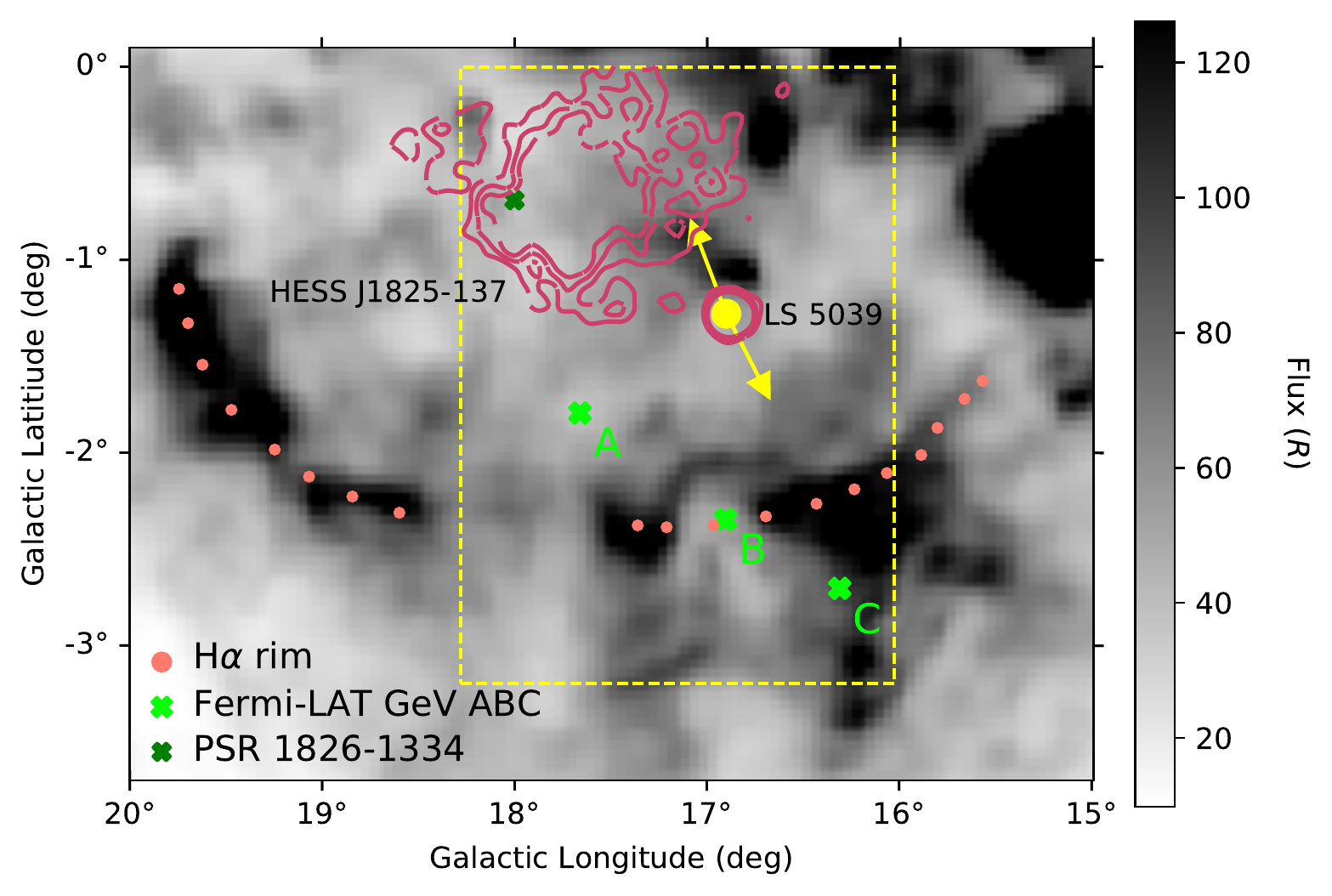}
    \caption{H$\alpha$ emission from the FWHM survey towards \HESSsource and surrounding regions \citep{2003ApJS..146..407F}. \HESSsource can be seen by $\sigma=1$, $2$ and $3$ purple TeV contours with PSR\,J1826-1334 represented by the dark green cross and LS-5039 by the yellow dot lying to the lower right with yellow radio jets described by \protect\cite{2002A&A...393L..99P}. GeV regions GeV-A, GeV-B and GeV-C can be seen as green neon crosses. The H$\alpha$ rims noted by \protect\cite{10.1111/j.1365-2966.2008.13761.x} and \protect\cite{2016MNRAS.458.2813V} are shown by pink dots. A closer look at the region contained within yellow box is shown in Fig.\,\ref{fig:H_alpha_zoomed}.}
    \label{fig:H_alpha}
\end{figure}

\begin{figure}
    \centering
    \includegraphics[width=\columnwidth]{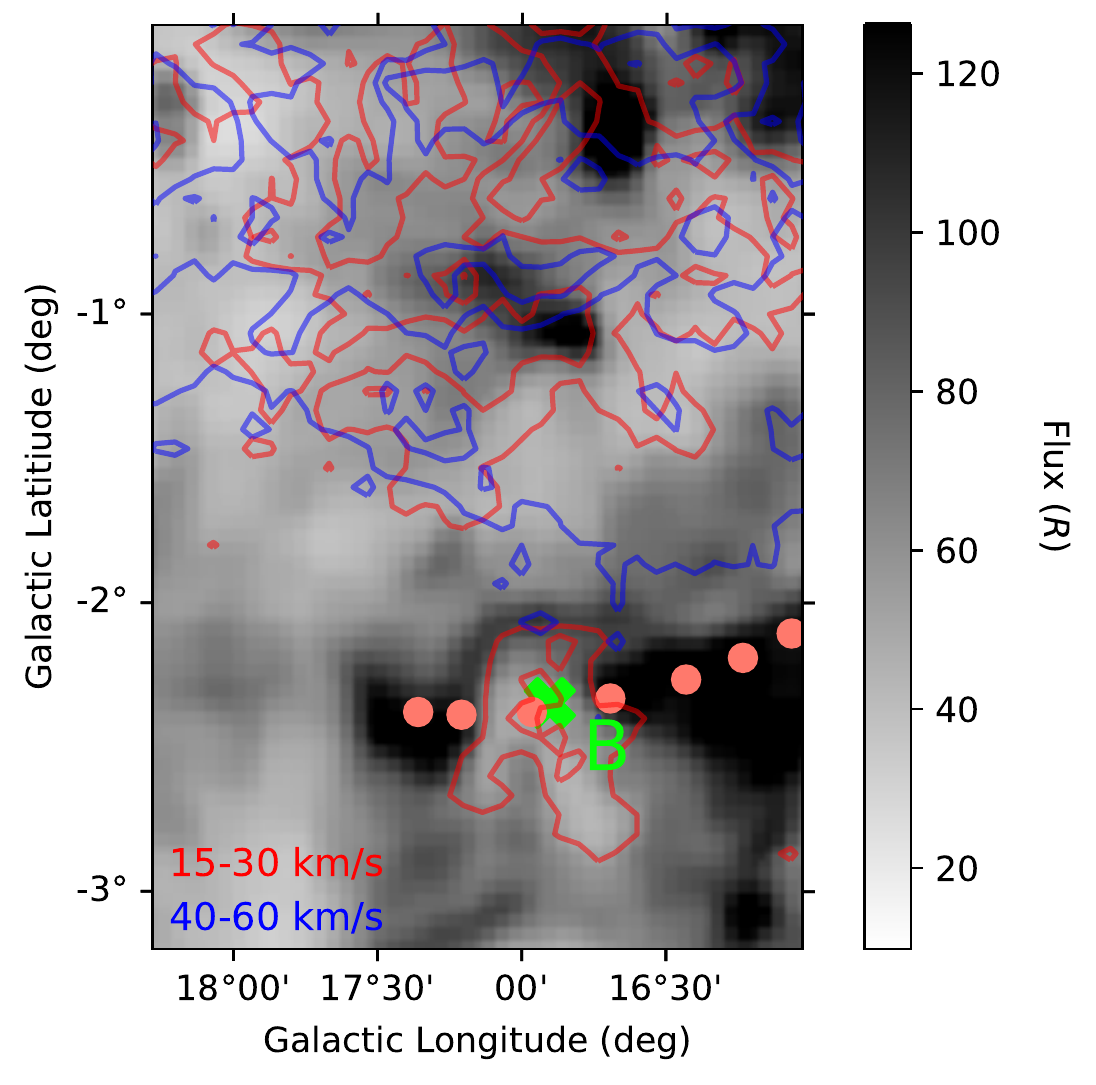}
    \caption{Zoomed in H$\alpha$ intensity overlayed with Nanten 12CO(1-0) in the $15-30~\kmpersec$ (red) and $40-60~\kmpersec$ (blue) range. The H$\alpha$ rims are indicated by the pink dots. Note that this corresponds to the yellow box in Fig.\,\ref{fig:H_alpha}. The dense CO cloud in the $15-30~\kmpersec$ velocity range towards GeV-B can be seen to anti-correlate with the H$\alpha$ emission.}
    \label{fig:H_alpha_zoomed}
\end{figure}

An intensity map of H$\alpha$ emission towards \HESSsource and surrounding regions can be seen in Fig.\,\ref{fig:H_alpha} from the FWHM survey \citep{2003ApJS..146..407F}. The H$\alpha$ rims detected by \cite{10.1111/j.1365-2966.2008.13761.x} and \citep{2016MNRAS.458.2813V} can clearly be seen and are located $\approx 120~\pc$ from PSR\,J1826-1334 if it lies at the same distance ($3.9~\kpc$) as the pulsar. From hydrodynamical simulations, the supernova remnant radius is, at least, four times the radius of this PWN \citep{2001ApJ...555L..49V}; this suggests a SNR radius of $140~\pc$ as calculated by \cite{2016MNRAS.458.2813V} agreeing with the rim of ionized gas seen in Fig.\,\ref{fig:H_alpha} and predictions made by \cite{2009ASSL..357..451D}.
\par 
\par 
Overlaying combined molecular and HI contours onto the H$\alpha$ map, (see Fig.\,\ref{fig:H_alpha_zoomed}), it can be seen the CO(1-0) cloud in the $15-30~\kmpersec$ range noted in section \ref{sec:CO_data} overlaps a region of reduced H$\alpha$ emission. This may indicate that the cloud is in the foreground or that the CO(1-0) cloud is surrounded by H$\alpha$ gas.
\par 
Two different methods were utilised to calculate the density of ionised hydrogen towards the regions of interest. The details of these calculations are provided in appendix \ref{sec:Halpha_method}. The results of both methods are shown in Table\,\ref{tab:Halpha_density}. Method A assumes that the density of photons is approximately equal to the density of ionised gas, assuming that atoms are not re-excited by an external source. Method B considers basic radiation transfer. It is expected that the ratio of ionised to neutral hydrogen atoms is $\approx 10^{-6}$ which agrees with both methods \citep{2011piim.book.....D}. Therefore ionised hydrogen does not significantly contribute to the total density of the ISM.

\section{Particle Transport} \label{sec:particle_transport}
\begin{figure}
    \centering
    \includegraphics[width=\columnwidth]{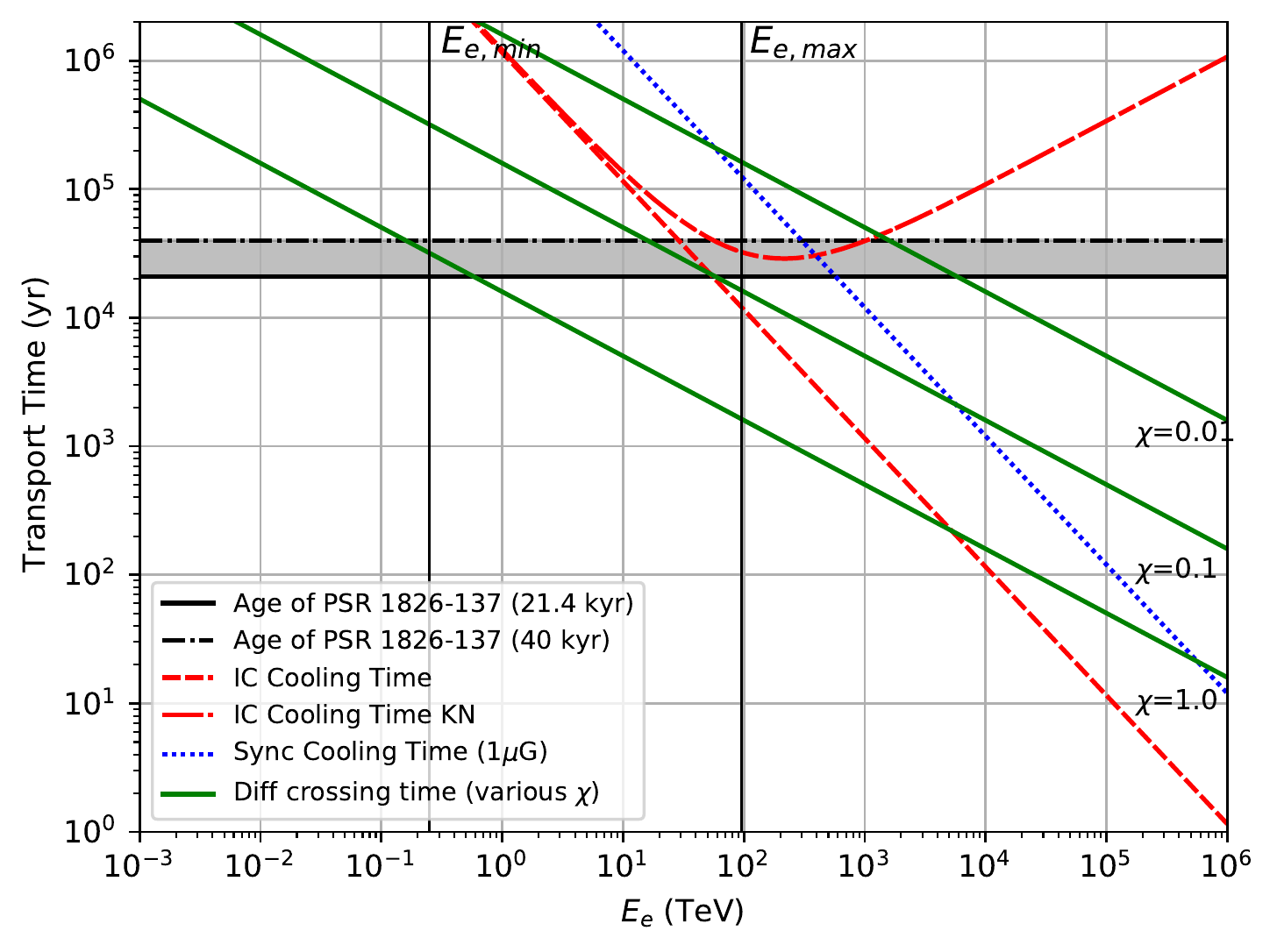}
    \caption{Transport time for particles to traverse from PSR\,1826-1336 to GeV-B versus cooling time of synchrotron and IC processes. The ambient density of the ISM is assumed to be $n=1~\centimeterminusthree$ .The black horizontal solid and dashed lines show the two possible ages of PSR\,J1826-1334 ($t=21.4~\si{yr}$ and $t=40~\si{yr}$ respectively), the dotted blue line shows the cooling time through synchrotron losses at $1~\si{\micro G}$ while the two red dashed lines is through IC losses in the Thompson and Klein Nishina regime. The solid lines with varying $\chi$ values shows the values necessary for particles with that energy to reach GeV-B in the available time through ISM with magnetic field $B=1~\si{\micro G}$. The inferred minimum and maximum electron energy, $E_{\text{e, min}}$ and $E_{\text{e, max}}$, emitted by the pulsar wind nebula is shown by the vertical solid lines.}
    \label{fig:Particle_Transport}
\end{figure}
After having mapped out the ISM, we can now consider the diffusive transport of high energy particles. In this study the spectral energy distribution modelling assumes that the high energy particles are able to enter GeV-ABC in sufficient number and energy range to produce the $\GeV$ gamma radiation. This section will look into the validity of this assumption assuming a purely diffusive scenario and looking at the cooling time of particles and how it affects the particle transport.
\par
Once high energy particles are emitted by the PWN (or progenitor SNR) they must traverse the interstellar medium before entering the region towards GeV-ABC. In a purely diffusive scenario, the distance that particles of energy $E$ diffuse into the ISM in time $t$ is estimated by

\begin{align}
    R\qty(E,t)&=\sqrt{2D\qty(E,B)t}\quad \qty[\si{\centi\meter}] \label{eq:R_diffusion}
\end{align}

where

\begin{align}
    D\qty(E,B)&=\chi D_0\sqrt{\frac{E/\TeV}{B/3~\si{\micro G}}}\quad [\si{\centi\meter\squared\per\second}] \label{eq:Diffusion}
\end{align}

$D_0=1\times 10^{29}\si{\centi\meter\squared\per\second}$ is the galactic diffusion coefficient at $1~\TeV$ and $\chi$ takes values of around $0.01$ (with variation) \citep{1990acr..book.....B,2007Ap&SS.309..365G}. As particles traverse the ISM they suffer energy losses through IC, bremsstrahlung and synchrotron radiation. The cooling time for bremsstrahlung, $t_\text{brem}$, Inverse Compton, $t_\text{IC}$, and synchrotron, $t_\text{sync}$, loss processes is given by:

\begin{subequations}
    \begin{align}
        t_\text{brem}& \approx \frac{4\times 10^7}{n~\centimeterminusthree}\quad\si{yr} \\
        t_\text{IC}& \approx \begin{cases}
             {3\times 10^5} U^{-1} (\frac{E}{\TeV})^{-1}\quad\si{yr} &\text{Thompson Regime} \\
             3.1\times 10^5 U^{-1} (\frac{E}{\TeV})^{-1}f_\text{KN}^{-1}\quad\si{yr} &\text{KN Regime}
        \end{cases} \\
        t_\text{sync}& \approx 12\times 10^6 \qty(\frac{B}{\si{\micro G}})^{-2} \qty(\frac{E}{\TeV})^{-1}\quad\si{yr} \label{eq:cooling_time_sync}
    \end{align}
\end{subequations} \label{eq:cooling_times}

where $U=0.26~\si{\electronvolt\per\centi\meter\cubed}$ is the energy density of the cosmic microwave background and $f_\text{KN}$ is the Klein Nishina (KN) suppression factor given by \cite{2005MNRAS.363..954M}:

\begin{align}
    f_\text{KN}&=\qty(1+40\frac{E}{\TeV} kT_{\eV})^{-1.5}
\end{align}

\noindent for an electron with energy $E$ interacting with photon field with temperature $T$ (with $kT$  in units of $\eV$). If the density of the ISM is $n=1$, the time it takes for particles of varying energies to be emitted by the PWN and travel to GeV-B ($\approx 70~\pc$) and the cooling time is shown in Fig.\,\ref{fig:Particle_Transport}. The intersection of the diffusion time and the age of PSR\,J1826-1334 represents the minimum particle energy that can reach GeV-B. Naturally if the pulsar is older, more lower energy particles can reach GeV-B. The maximum energy of electrons able to reach GeV-B is found through the intersection of the diffusion line and the IC cooling time (the quickest process where electrons lose most of their energy).
\par 
In IC processes, the final energy of the photon, $E_{\Gamma, \TeV}$, is related to initial electron energy, $E_{e,\TeV}$ and initial photon energy $E_{i,\eV}$ through:
\begin{align}
    E_{\Gamma,\TeV}&=E_{e,\TeV} \frac{h}{\qty(1+h^{4/5})^{5/4}}
\end{align}


where $h\approx 31.5E_{e,\TeV}E_{i,\eV}$ \citep{2009ARA&A..47..523H}. Photons up to $54~\TeV$ has been observed towards \HESSsource \citep{2019A&A...621A.116H}. Assuming IC interactions with the cosmic microwave background are responsible for this emission, this is equivalent to an electron with maximum energy of $96~\TeV$. Similarly photons as low as $1.26~\GeV$ has been observed by \fermi-LAT \citep{2020A&A...640A..76P}, which is equivalent to minimum electron energy of $0.25~\TeV$. This electron energy range is shown by the vertical lines in figure \ref{fig:Particle_Transport}. This further limits how many electrons are able to diffuse to GeV-B.
\par
The region around the PWN can harbour a strong magnetic field strength compared the surrounding ISM. Eq.\,\ref{eq:cooling_time_sync} outlines the cooling time for the synchrotron processes as a function of electron energy and magnetic field; (as shown by Fig.\,\ref{fig:Particle_Transport}). Consequently electrons in the zone around the pulsar wind nebula will experience stronger synchrotron losses compared to what is shown in Fig.\,\ref{fig:Particle_Transport}.

\section{SED modelling of the gamma-ray emission}

Two main pathways are possible for the production of $\GeV$ gamma radiation. In a hadronic scenario, proton-proton interactions with the ISM leading to the production of neutral pions which, in turn, decay into gamma-radiation. Leptonic scenarios include synchrotron emission associated with the magnetic field pervading in the ISM, inverse-compton emission with the cosmic microwave background and Bremsstrahlung interactions with the ISM.
\par 
Two types of particle accelerators will be considered; continuous and impulsive accelerators. Continuous accelerators constantly inject particles into the interstellar medium throughout their lifetime. For this study, it will be assumed that particles will be injected at a constant energy rate. Continuous accelerators may include pulsars and stellar clusters for example. On the other hand an impulsive accelerator, such as a supernova remnant, injected particles in one big burst in the past.
\par 
The particles that are injected are then free to undergo interactions producing radio to gamma-ray emission. The following section will describe potential particle accelerators that may result in the GeV gamma radiation as seen by \fermi-LAT. The model utilised in this study takes the initial particle spectrum and then lets the system evolve over the age of the particle accelerator. After the allocated time has passed, the final particle spectrum is calculated and the gamma-ray spectrum is extracted. For further explanation of the process utilised in this study, please refer to Appendix \ref{sec:newsedprod}.
\par 
The ISM density of the region to be modelled will utilise the data calculated in section \ref{sec:gas}. In turn the magnetic field strength, which affects the production of synchrotron radiation, is related to the density of ISM through the relation \citep{Crutcher_1999}:
\begin{align}
    B\qty(n_H)&=100\sqrt{\frac{n_H}{10^4~\centimeterminusthree}}~\si{\micro G}
\end{align}

Note that the updated version of this relation provides a slightly higher magnetic field estimation \citep{2010ApJ...725..466C}. Crutcher's relation computes the maximum magnetic field in a molecular cloud, allowing estimations calculated using \citep{Crutcher_1999} to be acceptable for this study.

\subsection{Potential Particle Accelerators}

\subsubsection{\HESSsource (PWN - Continuous)}

As shown in Fig.\,\ref{fig:CO_data}, it appears the Fermi GeV-ABC might be an extension of HESS\,J1825-137.
\par 
A part of the spin down power of HESS\,J1825-137, $2.8\times 10^{36}~\ergspersecond$, is channelled into accelerating particles that propagate out of the system. It was found that the major axis of gamma-emission is to the south-west of the pulsar towards GeV-ABC \citep{2019A&A...621A.116H}. The asymmetry in the gamma-ray emission may indicate an asymmetry in the particle emission by the PWN. The PWN would be a continuous source of high energy electrons towards the new region of GeV \fermi-LAT emission.

\subsubsection{\HESSsource Progenitor (SNR - Impulsive)} \label{sec:sed_modelling_snr}

Here we assume the progenitor SNR is an impulsive accelerator where the bulk of the cosmic rays escape the system very early and travel ahead of the SNR. Cosmic rays of energies $E$ escape the SNR in time $\chi$:

\begin{align}
    \chi(E)&=t_\text{Sedov}\qty(\frac{E}{E_\text{max}})^{-1/\delta}
\end{align}

\noindent where $t_\text{Sedov}=200~\si{yr}$ is the onset of the Sedov Phase of a SNR, $\delta=2.48$ is a parameter describing the energy dependent release of cosmic rays and $E_\text{max}=500~\TeV$ is the maximum possible cosmic ray proton energy \citep{2009MNRAS.396.1629G}. The $\TeV$ cosmic rays responsible for the emission of gamma-rays towards GeV-ABC have an escape time of $\sim 2~\kiloyear$. This is negligible compared to age of the pulsar ($21.4~\kiloyear$). The size of the SNR during the Sedov phase can be determined through \citep{2008ARA&A..46...89R}:

\begin{align}
    R&=0.31E_{51}^{1/5}\qty(\mu_1/1.4)^{-1/5}n^{-1/5}t_\text{yr}^{2/5}~\pc
\end{align}

where $E_{51}$ is the kinetic energy of the SNR in units of $10^{51}~\ergs$, $\mu_1$ is the mean mass per particle and $n$ is the background ISM density. If we assume $E_{51}=1$, $n=1~\centimeterminusthree$ and $\mu=1.41$; at age $2~\kiloyear$, the SNR will have a radius of $\sim 7~\pc$. The $\TeV$ cosmic-rays will escape the SNR at this radius and diffuse ahead of the SNR to GeV-ABC. Therefore the SNR progenitor associated with HESS\,J1825-137, as noted by \cite{10.1111/j.1365-2966.2008.13761.x} and \cite{ 2016MNRAS.458.2813V}, can be approximated as an impulsive source of high energy particles.  Additionally, it is generally believed that $10-30\%$ of the $10^{51}~\ergs$ of kinetic energy released in a supernova is channelled into accelerated high energy particles by the subsequent supernova remnant.
\par 
The distance to the PWN and SNR associated with \HESSsource will be assumed to be $3.9~\kpc$ \citep{1993ApJ...411..674T}. For this reason ISM parameters in the $40-60~\kmpersec$ velocity range (see Table. \ref{tab:total_density}) will be used in the SED modelling of GeV-A, GeV-B and GeV-C. Due to the anti-correlation of CO(1-0) to \halpha emission as seen in Fig.\,\ref{fig:H_alpha_zoomed}, the dense gas towards GeV-B in the $15-30~\kmpersec$ velocity will also be considered as a target for high energy particles to emit radiation. Due to individual gas motion compared to the galactic rotation curve \citep{1993A&A...275...67B}, this region of dense gas may be located at the same distance as \mbox{HESS\,J1825-137}.

\subsubsection{LS\,5039 (accretion powered - Continuous)}
\label{sec:LS_5039_cotinuous_proposal}

LS\,5039 is a microquasar and X-ray binary system \citep{1997A&A...323..853M}. LS\,5039 contains an O type star in orbit around an unknown compact object with mass $\approx3.7~M_\odot$ \citep{2005A&A...429..755P}. This high mass is greater than standard neutron star masses leading to the possibility of the compact object being a black hole. The high mass of the compact object suggests that the progenitor was born in the binary system with a mass greater than the O type star ($M_O=22.9~M_\odot$). The age of LS\,5039 is unknown; the lifetime of an O-type star is of order a few million years, giving an upper limit to the age of the system. The minimum and maximum plausible ages, of $1\times 10^3~\si{yrs}$ and $1\times10^6~\si{yrs}$ respectively, will be considered in the modelling \citep{2012A&A...543A..26M}. \cite{2012A&A...543A..26M} aimed to find the galactic trajectory of LS\,5039 to determine its birthplace. Depending on where LS\,5039 was born, \cite{2012A&A...543A..26M} gives the age of the system to be between $0.1-1.2~\si{Myr}$.  Therefore an age of $10^5~\si{yr}$ will also be considered in the SED modelling of this paper. In modelling the SED, these ages reflect the time when high energy particles enter GeV-ABC. Assuming diffusion is the particle transport method as in section\,\ref{sec:particle_transport}, the transportation time of high energy particles between LS\,5039 and GeV-ABC ($\approx 10^4~\si{yrs}$) is negligible compared to the age of LS\,5039.
\par 
After formation, the compact object associated with LS\,5039 continuously accretes matter from its star companion allowing particles to be accelerated in a relativistic radio jet.  This may be a continuous accelerator of particles to form the new $\GeV$ region as seen by \fermi-LAT. Radio jets described by \cite{2002A&A...393L..99P} can be seen in figure \ref{fig:H_alpha}. The average accretion luminosity of LS\,5039 was calculated by \cite{2005MNRAS.364..899C} to be $L_\text{acc}=8\times 10^{35}~\ergspersecond$. The luminosity released in the vicinity of LS\,5039 is given by: $L_{\text{radio, }0.1-100~\si{\giga\hertz}}\approx 1.3\times 10^{31}~\ergspersecond$ \citep{1998A&A...338L..71M}, $L_{\text{x-ray, }3-30\keV}=0.5-5\times 10^{34}~\ergspersecond$ \citep{2005ApJ...628..388B} and $L_{>100~\GeV}=2.7\times 10^{35}~\ergspersecond$ \citep{2005MNRAS.364..899C}.
Therefore \citep{2005MNRAS.364..899C} concluded that approximately one third of the accretion luminosity is channelled into the relativistic jets. The remaining $5.5\times 10^{35}~\ergspersecond$ can be channelled into GeV-ABC. Given the distance estimate to LS\,5039 of $2.5~\kpc$, the ISM within the $15-30~\kmpersec$ regime will be considered. It has been noticed that the radio jets are persistent with variability on day, week and year time scales \citep{2015MNRAS.451...59M}.
\par 

\subsubsection{LS 5039 Progenitor (SNR - Impulsive)}

Whether the compact object within LS\,5039 is a black hole or neutron star, the compact object is the result of a star gone supernova. By this logic an impulsive source of high energy particles occurred sometime in the past. At the time of writing, no clear SNR has been linked to LS\,5039. If LS\,5039 has age of order $10^5~\si{yrs}$, any SNR will be too old to be detected.
\par 

\subsection{Spectral Energy Distribution}

\begin{table*}
    \begin{threeparttable}
    \caption{SED model parameters matching the observed emission of GeV-A, B and C for a hadronic scenario. The particle accelerators considered are the impulsive progenitor SNR associated with PSR\,1826-1334 and the continuous accelerator associated with the pulsar wind nebula, HESS\,J1825-137. High energy particles are assumed to be injected with a power law spectra with an exponential cutoff: $\dv{N}{E} \propto E^{-\Gamma}\exp\qty[E/E_c]$.}
	\centering
	\begin{tabular}{c|cccccccc}
		\hline
		 Accelerator & \multicolumn{8}{c}{\rule{2cm}{0.5pt} Hadronic \rule{2cm}{0.5pt}}\\
		 PSR\,1826-1334 or SNR & Peak & $n_H$ ($\centimeterminusthree$) & \footnote{1}$W$ or \footnote{2}$\dot{W}$ & & $\Gamma$ & $E_C$ ($\TeV$) & \footnote{3}$W_\text{SNR}$ or \footnote{4}$\dot{W}_\text{tot}$  & \\
		 \hline
		 Impulsive (SNR) & A & $14$ & $1.0\times 10^{50}$ & $\ergs$ & $2.0$ & $50$ & $6.0\times 10^{51}$ & $\ergs$ \\
		$t=21\times 10^3~\si{yrs}$ & \textbf{B} & $\boldsymbol{79.8}$ & $\boldsymbol{1.5\times 10^{49}}$ & & $\boldsymbol{2.0}$ & $\boldsymbol{50}$ & $\boldsymbol{4.7\times 10^{50}}$ & \\
		 & B & $7.0$ & $1.5\times 10^{50}$ & & $2.0$ & $50$ & $4.7\times 10^{51}$ & \\
		 & C & $1.0$ & $1.2\times 10^{51}$ & & $2.0$ & $50$ & $7.3\times 10^{52}$ & \\
		 \hline
		 Impulsive (SNR) & A & $14$ & $1.0\times 10^{50}$ & $\ergs$ & $2.0$ & $50$ & $6.1\times 10^{51}$ & $\ergs$ \\
		$t=40\times 10^3~\si{yrs}$ & \textbf{B} & $\boldsymbol{79.8}$ & $\boldsymbol{1.5\times 10^{49}}$ & & $\boldsymbol{2.0}$ & $\boldsymbol{50}$ & $\boldsymbol{4.7\times 10^{50}}$ & \\
		 & B & $7.0$ & $1.5\times 10^{50}$ & & $2.0$ & $50$ & $4.7\times 10^{51}$ & \\
		 & C & $1.0$ & $1.1\times 10^{51}$ & & $2.0$ & $50$ & $6.7\times 10^{52}$ & \\
		 \hline
		Continuous (PWN) & A & $14$ & $1.2\times 10^{38}$ & $\ergspersecond$ & $2.0$ & $50$ &  & \\
		$t=21\times 10^3~\si{yrs}$ & B & $79.8$ & $2.0\times 10^{37}$ & & $2.0$ & $50$ &  & \\
		 & B & $7.0$ & $2.2\times 10^{38}$ & & $2.0$ & $50$ &  & \\
		 & C & $1.0$ & $1.7\times 10^{39}$ & & $2.0$ & $50$ & $1.8-2.0\times10^{39}$ & $\ergspersecond$ \\
		 \hline
		Continuous (PWN) & A & $14$ & $8.0\times 10^{37}$ & $\ergspersecond$ & $2.0$ & $50$ & &  \\
		$t=40\times 10^3~\si{yrs}$ & B & $79.8$ & $1.25\times 10^{37}$ & & $2.0$ & $50$ & & \\
		 & B & $7.0$ & $1.25\times 10^{38}$ & & $2.0$ & $50$ & & \\
		 & C & $1.0$ & $8.5\times 10^{38}$ & & $2.0$ & $50$ & $0.9-1.0\times 10^{39}$ & $\ergspersecond$ \\
		\hline
	\end{tabular}
	\begin{tablenotes}
	\item \footnotemark[1]{$W$: Energy budget of high energy particles within individual clouds (see Fig. \ref{fig:CO_data})} 
    \item \footnotemark[2]{$\dot{W}$: Particle injection luminosity of high energy particles into individual clouds} 
    \item \footnotemark[3]{$W_\text{SNR}$: Injected energy budget of high energy particles within progenitor SNR (see equation \ref{eq:filling_factor})} 
	\item \footnotemark[4]{$\dot{W}_\text{tot}$: Total injection luminosity of all three regions by PWN}
    \item Plausible scenarios are shown in bold
    \item Matching scenarios have systematic variation of up to $56\%$ in energy budget $W$ or luminosity $\dot{W}$, $12\%$ in the spectral index $\Gamma$ and $12\%$ in the cutoff energy $E_c$ (see text and figure \ref{fig:example_SED})
    \end{tablenotes}
	\label{tab:SED_sum_1825_hadronic}
	\end{threeparttable}
\end{table*}

\begin{table*}
    \begin{threeparttable}
    \caption{Same as Table. \ref{tab:SED_sum_1825_hadronic} but parameters in a leptonic origin.}
	\centering
	\begin{tabular}{c|cccccccc}
		\hline
		Accelerator & \multicolumn{8}{c}{\rule{2cm}{0.5pt} Leptonic \rule{2cm}{0.5pt}}\\
		 PSR\,1826-1334 or SNR & Peak & $n_H$ ($\centimeterminusthree$) & $W$ or $\dot{W}$ & & $\Gamma$ & $E_C$ ($\TeV$) & $W_\text{SNR}$ or $\dot{W}_\text{tot}$ & \\
		 \hline
		 Impulsive (SNR) & A & $14$ & $1.2\times 10^{49}$ & $\ergs$ & $2.0$ & $10$ & $7.3\times 10^{50}$ & $\ergs$ \\
		$t=21\times 10^3~\si{yrs}$ & B & $79.8$ & $9.0\times 10^{48}$ & & $2.0$ & $30$ & $2.8\times 10^{50}$ & \\
		 & B & $7.0$ & $8.0\times 10^{48}$ & & $2.0$ & $30$ & $2.5\times 10^{50}$ & \\
		 & C & $1.0$ & $7.0\times 10^{48}$ & & $2.0$ & $10$ & $4.3\times 10^{50}$ & \\
		 \hline
		 Impulsive (SNR) & A & $14$ & $1.4\times 10^{49}$ & $\ergs$ & $2.0$ & $10$ & $8.5\times 10^{50}$ & $\ergs$ \\
		$t=40\times 10^3~\si{yrs}$ & B & $79.8$ & $6.0\times 10^{48}$ & & $1.0$ & $50$ & $1.9\times 10^{50}$ & \\
		 & B & $7.0$ & $3.0\times 10^{48}$ & & $1.5$ & $50$ & $9.4\times 10^{49}$ & \\
		 & C & $1.0$ & $7.6\times 10^{48}$ & & $2.0$ & $50$ & $4.6\times 10^{50}$ & \\
		 \hline
		 Continuous (PWN) & \textbf{A} & $\boldsymbol{14}$ & $\boldsymbol{1.5\times 10^{37}}$ & $\ergspersecond$ & $\boldsymbol{2.0}$ & $\boldsymbol{10}$ &  &  \\
		$t=21\times 10^3~\si{yrs}$ & \textbf{B} & $\boldsymbol{79.8}$ & $\boldsymbol{1.5\times 10^{37}}$ & & $\boldsymbol{2.0}$ & $\boldsymbol{10}$ &  & \\
		 & \textbf{B} & $\boldsymbol{7.0}$ & $\boldsymbol{1.5\times 10^{37}}$ & & $\boldsymbol{2.0}$ & $\boldsymbol{10}$ &  & \\
		 & \textbf{C} & $\boldsymbol{1.0}$ & $\boldsymbol{1.0\times 10^{37}}$ & & $\boldsymbol{2.0}$ & $\boldsymbol{10}$ & $\boldsymbol{4.0\times 10^{37}}$  & $\ergspersecond$ \\
		 \hline
		Continuous (PWN) & \textbf{A} & $\boldsymbol{14}$ & $\boldsymbol{1.0\times 10^{37}}$ & $\ergspersecond$ & $\boldsymbol{2.0}$ & $\boldsymbol{10}$ &  &  \\
		$t=40\times 10^3~\si{yrs}$ & \textbf{B} & $\boldsymbol{79.8}$ & $\boldsymbol{2.6\times 10^{36}}$ & & $\boldsymbol{1.7}$ & $\boldsymbol{10}$ &  & \\
		 & \textbf{B} & $\boldsymbol{7.0}$ & $\boldsymbol{1.6\times 10^{36}}$ & & $\boldsymbol{1.7}$ & $\boldsymbol{10}$ &  & \\
		 & \textbf{C} & $\boldsymbol{1.0}$ & $\boldsymbol{6.0\times 10^{36}}$ & & $\boldsymbol{2.0}$ & $\boldsymbol{10}$ & $\boldsymbol{1.8-1.9\times 10^{37}}$ & $\ergspersecond$ \\
		 \hline
	\end{tabular}
	\label{tab:SED_sum_1825_leptonic}
	\end{threeparttable}
\end{table*}

\begin{table*}
    \begin{threeparttable}
    \caption{Model parameters matching the observed emission of GeV-A, B and C for a hadronic scenario. The particle accelerators considered are the progenitor SNR associated with LS\,5039 (impulsive) or the accretion of matter by the companion star onto LS\,5039 (continuous). Example SED best fit is shown in Fig.\,\ref{fig:example_SED}. $W$ or $\dot{W}$ represents the energy budget. To see the regions used, refer to Fig.\,\ref{fig:HI_data}. The spectra of injected particles is represented by an exponential cutoff power law spectrum: $\dv{N}{E} \propto E^{-\Gamma}\exp\qty[E/E_c]$.}
	\centering
	\begin{tabular}{c|cccccccc}
		\hline
    	Accelerator & \multicolumn{8}{c}{\rule{2cm}{0.5pt} Hadronic \rule{2cm}{0.5pt}}\\
		 LS\,5039 or SNR & Peak & $n_H$ ($\centimeterminusthree$) & $W$ or $\dot{W}$ & & $\Gamma$ & $E_C$ ($\TeV$) & \footnote{5}$W_\text{SNR}$ or \footnote{6}$\dot{W}_\text{total}$ & \\
		 \hline
		 Impulsive (SNR) & A & $7.0$ & $8.0\times10^{49}$ & $\ergs$ & $2.0$ & $50$ & $3.9\times10^{51}$ & $\ergs$ \\
		 $t=1\times 10^3~\si{yrs}$ & B & $79.8$ & $6.0\times10^{48}$ & & $2.0$ & $50$ & $1.5\times10^{50}$ \\
		 & C & $3.6$ & $1.0\times10^{50}$ & & $2.0$ & $50$ & $4.9\times10^{51}$  \\
		 \hline
		 Impulsive (SNR) & A & $7.0$ & $1.0\times10^{50}$ & $\ergs$ & $2.0$ & $50$ & $4.9\times10^{51}$ & $\ergs$ \\
		 $t=1\times 10^5~\si{yrs}$ & B & $79.8$ & $8.0\times10^{48}$ & & $2.0$ & $50$ & $2.0\times10^{50}$ \\
		 & C & $3.6$ & $1.5\times10^{50}$ & & $2.0$ & $50$ & $7.4\times10^{51}$ \\
		 \hline
		 Impulsive (SNR) & A & $7.0$ & $3.0\times10^{51}$ & $\ergs$ & $1.0$ & $50$ & $1.5\times10^{53}$ & $\ergs$ \\
		 $t=1\times 10^6~\si{yrs}$ & B & $79.8$ & $4.0\times10^{49}$ & & $1.5$ & $50$ & $1.0\times10^{51}$ \\
		 & C & $3.6$ & $4.0\times10^{51}$ & & $1.0$ & $50$ & $2.0\times10^{53}$ \\
		 \hline
		Continuous (accretion) & A & $7.0$ & $2.3\times10^{39}$ & $\ergspersecond$ & $2.0$ & $50$ \\
		$t=1\times 10^3~\si{yrs}$ & B & $79.8$ & $2.0\times10^{38}$ & & $2.0$  & $50$ & &  \\
		 & C & $3.6$ & $4.5\times10^{39}$ & & $2.0$ & $50$ & $7.0\times 10^{39}$ & $\ergspersecond$ \\
		\hline
		Continuous (accretion) & A & $7.0$ & $2.8\times10^{37}$ & $\ergspersecond$ & $2.0$ & $50$ \\
		$t=1\times 10^5~\si{yrs}$ & B & $79.8$ & $2.0\times10^{36}$ & & $2.0$ & $50$ & & \\
		 & C & $3.6$ & $4.0\times10^{37}$ & & $2.0$ & $50$ & $7.0\times 10^{37}$ & $\ergspersecond$ \\ 
		\hline
		Continuous (accretion) & A & $7.0$ & $2.5\times10^{36}$ & $\ergspersecond$ & $2.0$ & $50$ \\
		$t=1\times 10^6~\si{yrs}$ & B & $79.8$ & $3.5\times10^{35}$ & & $2.0$ & $50$ \\
		 & C & $3.6$ & $4.0\times10^{36}$ & & $2.0$ & $50$ & $6.9\times 10^{36}$ & $\ergspersecond$ \\
		\hline
	\end{tabular}
	\begin{tablenotes}
	\item \footnotemark[5]{${W}_\text{SNR}$: Inferred energy budget of high energy particles inside progenitor SNR related to LS\,5039}
	\item \footnotemark[6]{$\dot{W}_\text{total}$: The total injection luminosity into all three clouds}
	\item Plausible scenarios are shown in bold
	\item Matching scenarios have systematic variation of up to $56\%$ in energy budget $W$ or luminosity $\dot{W}$, $12\%$ in the spectral index $\Gamma$ and $12\%$ in the cutoff energy $E_c$ (see text and figure \ref{fig:example_SED})
	\end{tablenotes}
	\label{tab:SED_sum_5039_hadronic}
	\end{threeparttable}
\end{table*}

\begin{table*}
    \begin{threeparttable}
    \caption{Same as Table.\,\ref{tab:SED_sum_5039_hadronic} but parameters in a leptonic scenario for gamma-ray emission.}
	\centering
	\begin{tabular}{c|cccccccc}
		\hline
		Accelerator & \multicolumn{8}{c}{\rule{2cm}{0.5pt} Leptonic \rule{2cm}{0.5pt}}\\
		LS\,5039 or SNR & Peak & $n_H$ ($\centimeterminusthree$) & $W$ or $\dot{W}$ & & $\Gamma$ & $E_C$ ($\TeV$) & $W_\text{SNR}$ or $\dot{W}_\text{total}$ \\
		 \hline
		 Impulsive (SNR) & A & $7.0$ & $4.0\times10^{48}$ & $\ergs$ & $2.0$ & $10$ & $2.0\times10^{50}$ & $\ergs$ \\
		 $t=1\times 10^3~\si{yrs}$ & B & $79.8$ & $2.5\times10^{48}$ & & $2.0$ & $10$ & $6.4\times10^{49}$ \\
		 & C & $3.6$ & $3.0\times10^{48}$ & & $2.0$ & $10$ & $1.5\times10^{50}$ \\
		 \hline
		 Impulsive (SNR) & A & $7.0$ & $5.0\times10^{48}$ & $\ergs$ & $1.7$ & $50$ & $2.5\times10^{50}$ & $\ergs$ \\
		 $t=1\times 10^5~\si{yrs}$ & B & $79.8$ & $2.0\times10^{49}$ & & $1.0$ & $10$ & $5.1\times10^{50}$ \\
		 & C & $3.6$ & $5.0\times10^{48}$ & & $2.0$ & $10$ & $2.5\times10^{50}$ \\
		 \hline
		 Impulsive (SNR) & A & $7.0$ & $3.0\times10^{50}$ & $\ergs$ & $1.7$ & $100$ & $1.5\times10^{52}$ & $\ergs$ \\
		 $t=1\times 10^6~\si{yrs}$ & B & $79.8$ & & & & \\
		 & C & $3.6$ & $2.0\times10^{50}$ & & $2.0$ & $10$ & $9.9\times10^{51}$ \\
		 \hline
		Continuous (accretion) & A & $7.0$ & $1.0\times10^{38}$ & $\ergspersecond$ & $2.0$ & $10$ \\
		$t=1\times 10^3~\si{yrs}$ & B & $79.8$ & $8.5\times10^{37}$ & & $2.0$ & $10$ & & \\
		 & C & $3.6$ & $5.0\times10^{38}$ & & $2.0$ & $10$ & $6.9\times 10^{38}$ & $\ergspersecond$ \\
		\hline
		Continuous (accretion) & A & $7.0$ & $3.5\times10^{35}$ & $\ergspersecond$ & $1.7$ & $10$ \\
		$t=1\times 10^5~\si{yrs}$ & B & $79.8$ & $9.0\times10^{35}$ & & $2.0$ & $10$ & & \\
		 & C & $3.6$ & $2.0\times10^{35}$ & & $1.5$ & $10$ & $1.5\times 10^{36}$ & $\ergspersecond$ \\
		\hline
		Continuous (accretion) & \textbf{A} & $\boldsymbol{7.0}$ & $\boldsymbol{2.0\times10^{35}}$ & $\ergspersecond$ & $\boldsymbol{1.7}$ & $\boldsymbol{10}$ \\
		$\boldsymbol{t=1\times 10^6~\si{yrs}}$ & \textbf{B} & $\boldsymbol{79.8}$ & $\boldsymbol{7.5\times10^{35}}$ & & $\boldsymbol{1.8}$ & $\boldsymbol{30}$ & & \\
		 & \textbf{C} & $\boldsymbol{3.6}$ & $\boldsymbol{9.0\times10^{34}}$ & & $\boldsymbol{1.5}$ & $\boldsymbol{10}$ & $\boldsymbol{1.0\times 10^{36}}$ & $\ergspersecond$ \\
		\hline
	\end{tabular}
	\label{tab:SED_sum_5039_leptonic}
	\end{threeparttable}
\end{table*}

\begin{figure}
    \centering
    \includegraphics[width=\columnwidth]{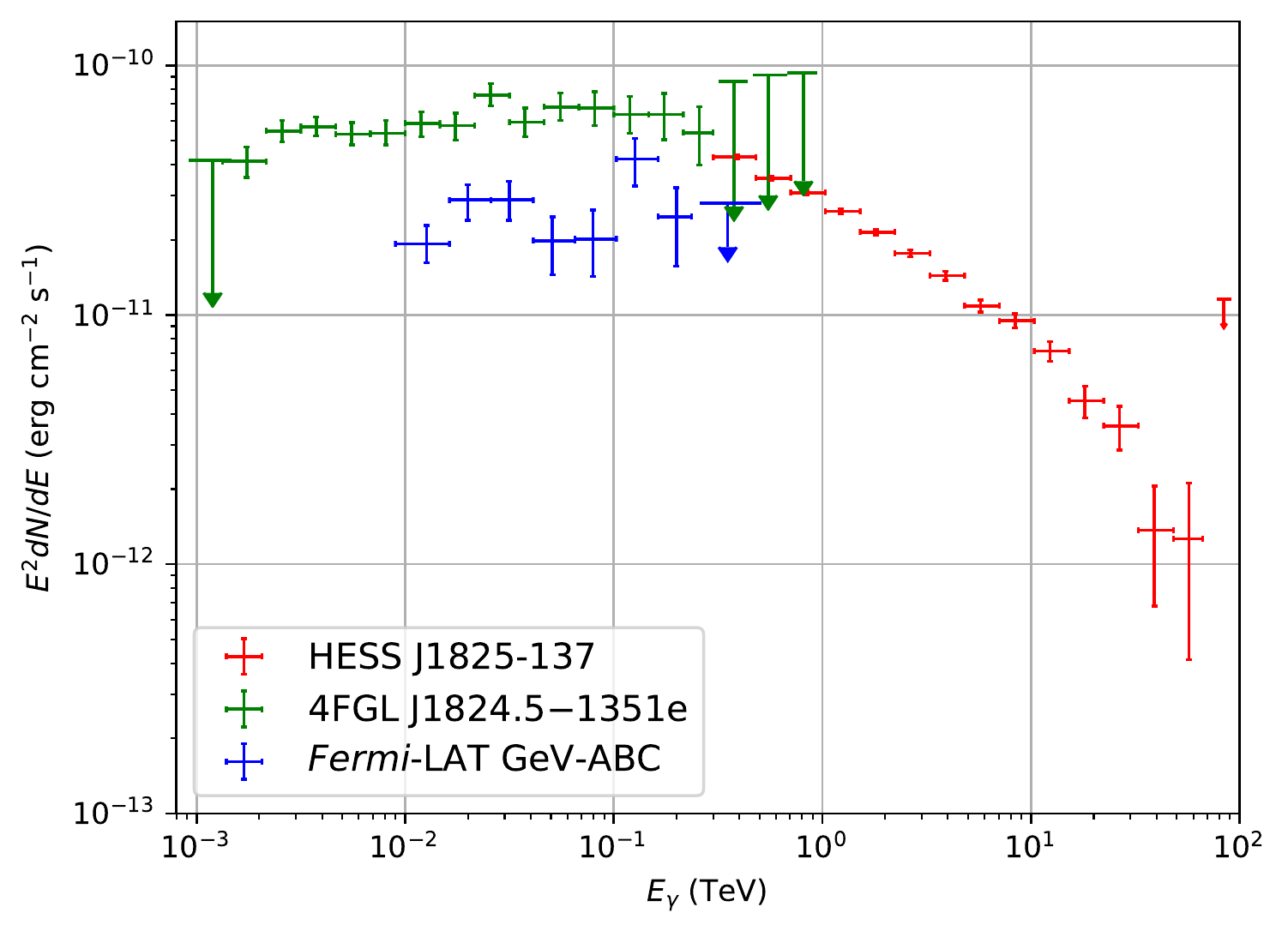}
    \caption{Spectral Energy Distribution of GeV-ABC as revealed by \protect\cite{Araya:2019dez} is shown in blue. The flux points of \HESSsource are represented by red. \protect\cite{2019A&A...621A.116H}. \textit{Fermi}-LAT data towards \HESSsource \citep{2020A&A...640A..76P} shown in green can be seen to follow the HESS data points as noticed by \protect\cite{2019A&A...621A.116H}.}
    \label{fig:Yama_Spectrum}
\end{figure}

\begin{figure*}
	\begin{subfigure}[b]{\columnwidth}
        \includegraphics[width=\textwidth]{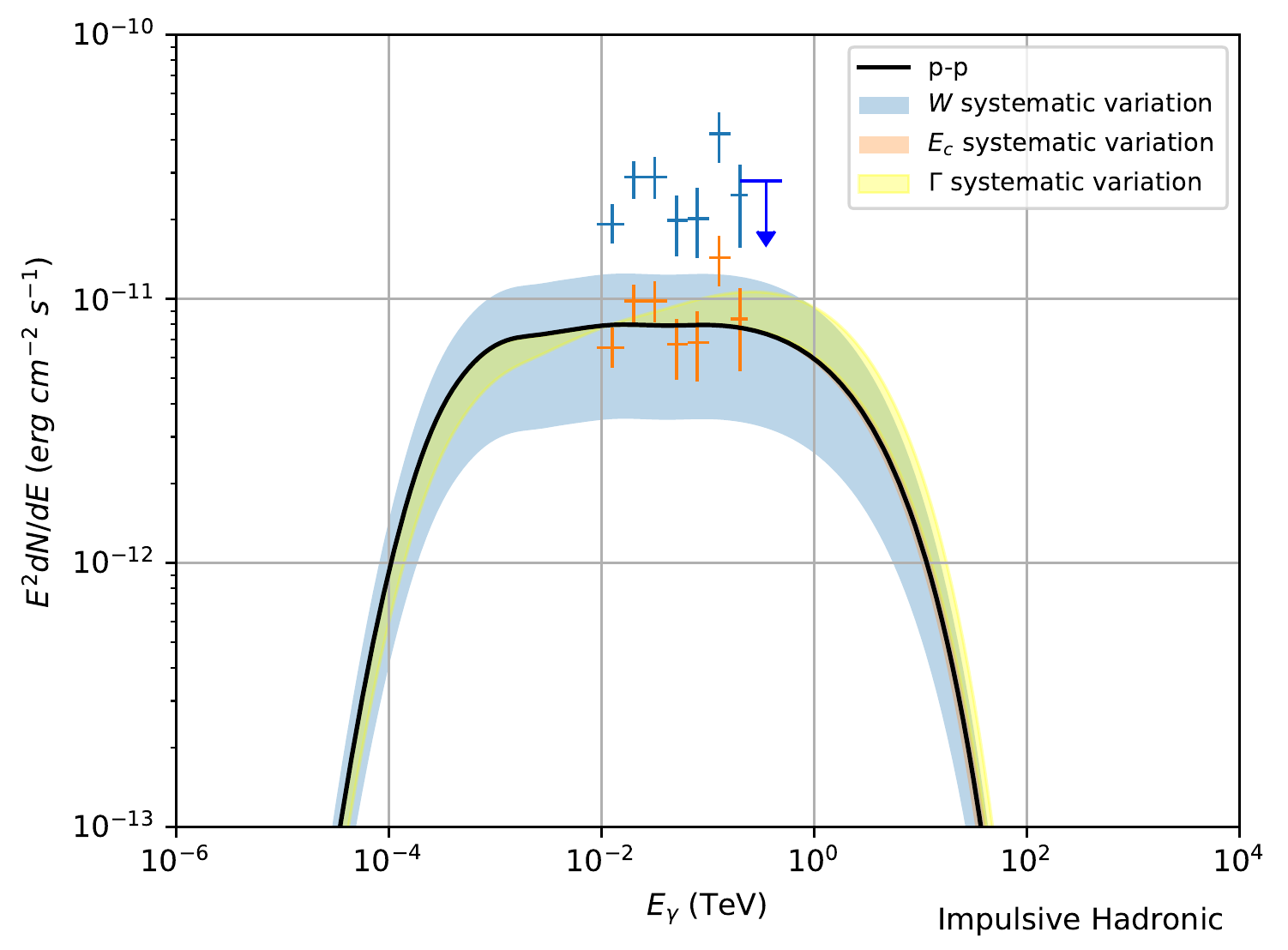}
	\end{subfigure}
	 ~
	\begin{subfigure}[b]{\columnwidth}
	    \includegraphics[width=\textwidth]{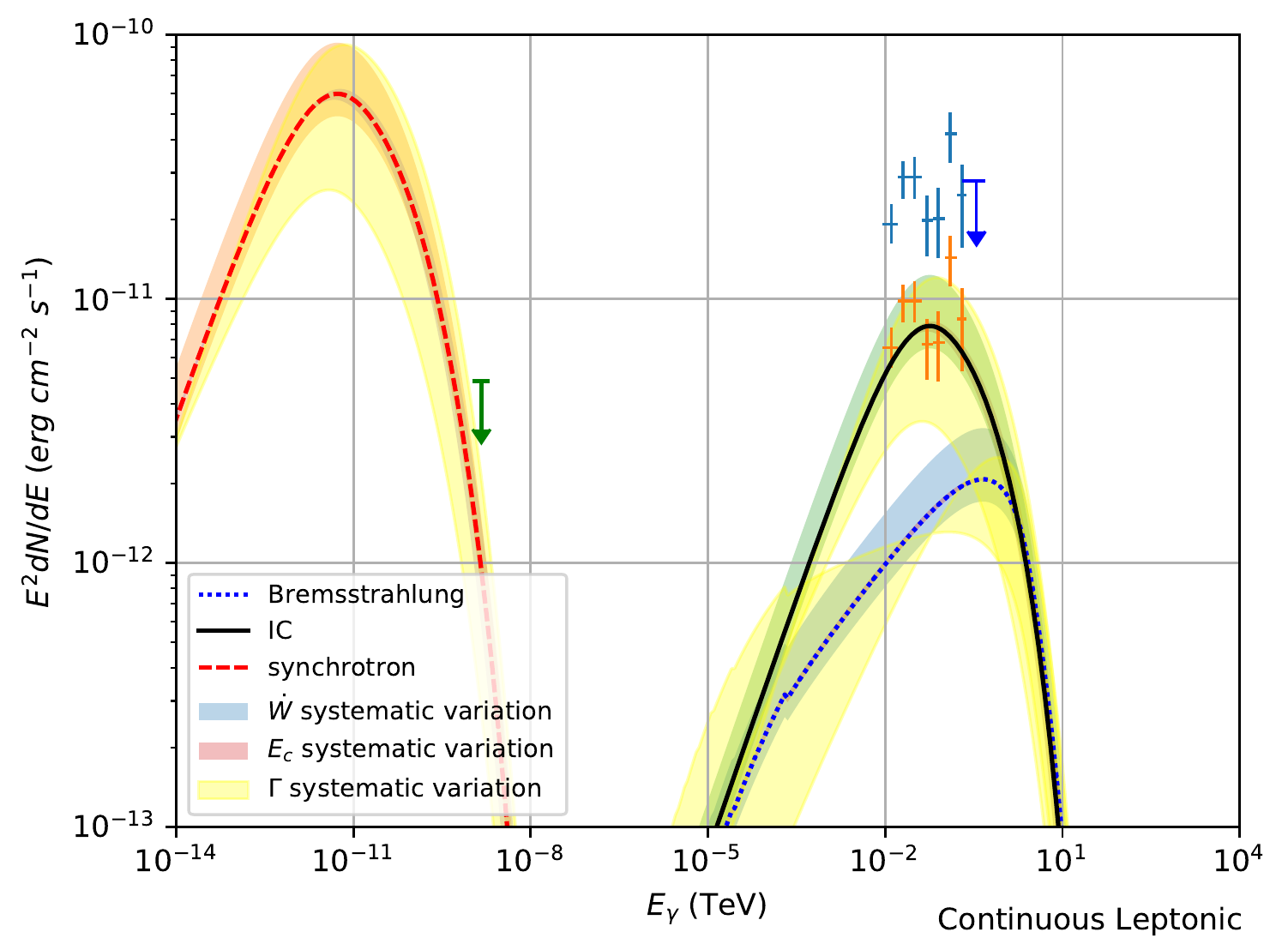}
	\end{subfigure}
    \caption{SED example for GeV-B with impulsive hadronic SED (\textit{left}) and continuous leptonic SED (\textit{right}). This assumes that \HESSsource (at age $40~\si{kyrs}$) is the source of acceleration. The upper data points represent the total SED as measured by \fermi-LAT. The lower data points is $34.0\%$ of this spectra due to source being only GeV-B. The green arrow is the ROSAT X-ray upper limit towards GeV-ABC. The blue, pink and yellow shaded regions represent the systematic variation of energy budget ($W$) or injection luminosity ($\dot{W}$), cutoff energy and spectral index respectively. See Tables\,\ref{tab:SED_sum_1825_hadronic} and \ref{tab:SED_sum_1825_leptonic} for input parameters.}
    \label{fig:example_SED}
\end{figure*}

The spectral analysis conducted by \cite{Araya:2019dez} towards GeV-ABC is shown in Fig.\,\ref{fig:Yama_Spectrum}. They found that a simple power-law ($\dv{N}{E}\propto E^{-\Gamma}$) best describes the spectrum with spectral index of $\Gamma = 1.92\pm 0.07_\text{stat}\pm 0.05_\text{sys}$ and integrated flux of $\phi_0=\qty(1.46\pm0.11_\text{stat}\pm0.13_\text{sys})\times10^{-9}~\si{photons\per\centi\meter\squared\per\second}$.
\par 
In the study by \cite{Araya:2019dez}, individual peaks GeV-A, B and C were found to have spectral indices  $\Gamma_A=1.78\pm0.25_\text{stat}$,  $\Gamma_B=1.7\pm0.4_\text{stat}$ and $\Gamma_C=1.43\pm 0.23_\text{stat}$ respectively. The extended GeV emission observed by \fermi-LAT will be modelled by approximating the spectra of GeV-ABC as coming from three sources corresponding to the peaks observed by \cite{Araya:2019dez}. By integrating the flux over all energy ranges for all three peaks and normalising to the spectra of GeV-ABC, the amount each peak contributes to the total flux can be determined. This assumes that the entirety of the GeV emission originates from the three peaks. As seen by Fig\,3 from \cite{Araya:2019dez} GeV\,A, B and C contains the majority of the GeV emission.. It was found that GeV-A, GeV-B and GeV-C, contributed $37\%$, $34\%$ and $29\%$ of the total $\GeV$ flux respectively. For each peak in the GeV gamma-ray emission region, spectral energy distributions based on different particles accelerators (e.g. PWN \HESSsource and LS\,5039) will be modelled and fit by eye to the data. Input parameters will also be varied to provide a range where the model matches the data. The ROSAT x-ray upper limit towards GeV-ABC was calculated using the ROSAT X-Ray Background Tool \citep{2019ascl.soft04001S}.
\par 
Input parameters of the spectral energy distribution modelling towards the new region of $\GeV$ emission can be seen in Tables\,\ref{tab:SED_sum_1825_hadronic} and \ref{tab:SED_sum_1825_leptonic} and Tables\,\ref{tab:SED_sum_5039_hadronic} and \ref{tab:SED_sum_5039_leptonic} for \HESSsource and LS\,5039 being the source of high energy particles respectively.
\par 
An example fit to the spectral energy distribution is shown in Fig.\,\ref{fig:example_SED} . It is assumed that both hadronic and leptonic particles followed an exponential cutoff power law injection spectra $\left(\dv{N}{E}\propto E^{-\Gamma}\exp\qty(-E/E_c)\right)$.  It is important to note that the energy budget/injection luminosity, $W$ or $\dot{W}$, that is inferred reflects the energy budget for each individual peak (GeV-A, GeV-B and GeV-C) and not the total energy budget/injection luminosity for the combination of all three regions.
\par 

Assuming constant cosmic ray density within a supernova remnant, the inferred energy ($W_\text{SNR}$) of the SNR can be calculated. The filling factor, $f_f$, is defined to be the ratio of the area of GeV-A, B or C to the projected area of the SNR. The inferred energy of the SNR is then given by:
\begin{align}
    W_\text{SNR}&=\frac{W}{f_f} \label{eq:filling_factor}
\end{align}

\subsubsection{HESS\,J1825-137 Progenitor}

A clear SNR rim can be seen in Fig.\,\ref{fig:H_alpha} connected to HESS\,J1825-137. The projected area of the SNR, with radius of $140~\pc$, is assumed to be $\approx 64\times 10^3~\si{\parsec^2}$. Note that the denser regions to the north of HESS\,J1825-137 shown in Fig\,\ref{fig:CO_data} and \ref{fig:HI_data} may dampen the northern expansion of the SNR associated with HESS J1825-137. This will affect the filling factor geometrically and in turn affect the inferred energy of the SNR as shown in Eq.\,\ref{eq:filling_factor}. If no particles have escaped, it is expected that $10^{50}~\ergs$ of energy remains within the SNR. As the SNR is definitely well into its Sedov phase, some cosmic rays will have escaped the system, lowering the remaining energy within the SNR.
\par 
For an individual model to be successful, it must allow sufficient energetics within all three clouds simultaneously. If the particle energetics impacting one cloud is too large, the model will be rejected. For this reason when looking at Tables.\,\ref{tab:SED_sum_1825_hadronic} and \ref{tab:SED_sum_1825_leptonic}, to determine if individual impulsive models were successful, the maximum energy budget/ injection luminosity will be compared to theoretical energetics.

\subsubsection{LS\,5039 Progenitor}

No clear SNR rim has been associated with LS\,5039. If the age of LS\,5039 is greater than $10^5~\si{yr}$ the SNR will have already dispersed into the The projected area of SNR will assume a minimum radius of $\approx80~\pc$ to completely encompass GeV-ABC as seen by \fermi-LAT. Equation\,\ref{eq:filling_factor} is then used to estimate the total energy of high energy particles remaining in the progenitor SNR of LS\,5039. The assumed minimum radius of SNR will lead to an underestimation of the inferred energy of the SNR associated with LS 5039.

\section{Discussion}
\par 
In this section we will discuss the results of the SED modelling and consider the possible accelerator scenarios. 

\subsection{Accelerator related to \HESSsource} \label{sec:1825_discussion}

Firstly we will examine the plausibility of an accelerator related to HESS\,J1825-137. The two sources of high energy particles are the progenitor SNR and PWN.

\subsubsection{A progenitor SNR (Impulsive)}

The progenitor SNR linked to \HESSsource is an impulsive accelerator; releasing $\approx10^{50}~\ergs$ of cosmic rays (with electrons making up $\approx 10^{48}~\ergs$) into the surrounding environment. The SNR expands and cosmic rays will escape from the system, decreasing the total energy of particles trapped inside the SNR. From spectral energy distribution modelling, the energy budget in regions GeV-A, B and C required to reproduce the SED of GeV $\gamma$-rays was obtained. The total SNR cosmic ray energy budget, $W_\text{SNR}$, is estimated by equation \ref{eq:filling_factor}.
\par 
To reproduce the SED of any of GeV-A, B or C requires the hadronic SNR energy budget to range between $5-730\times10^{50}~\ergs$, as shown in Table. \ref{tab:SED_sum_1825_hadronic}. It is possible that \HESSsource may be a possible hypernova candidate: supernova with kinetic energy greater than $10^{52}~\ergs$ \citep{2004ASSL..302..277N}. This is equivalent to a supernova releasing $10^{51}~\ergs$ of cosmic rays. A plausible scenario requires for all three GeV regions to simultaneously explain the gamma-ray spectrum. For both ages ($t=21$ and $40~\kiloyear$), only GeV-B has reasonable energetics assuming a higher ISM density; therefore a pure hadronic progenitor SNR scenario must be rejected unless a hypernova scenario is considered.
\par 
Hydrogen volume density is not constant across GeV-A, B and C. Equation \ref{eq:app_proton_proton_flux} shows that the spectra of gamma-rays from proton-proton interactions is proportional to the density of the target material. Assuming that the the high density cloud observed in the $15-30~\kmpersec$ velocity range lies at the same distance as HESS\,J1825-137 then GeV-B should appear brighter in gamma-rays compared to GeV-A and C assuming that the cosmic ray energy density over all three regions are constant. This is not the case, therefore the cosmic ray energy density in cloud A and C must be 7 and 80 times greater respectively than the energy density in cloud B. 
As discussed in section\,\ref{sec:sed_modelling_snr}, particles escape the SNR at age $\sim 2~\si{kyr}$ when it has a radius of $\sim 7~\pc$ and diffuse to GeV-ABC. By the time the particles have diffused the remaining distance to GeV-ABC ($\approx 130~\pc$) any local anisotropy at the GeV-ABC position will likely have been lost. Therefore an impulsive hadronic scenario cannot explain why GeV-A, B \& C have the same brightness.
\par 
A pure impulsive leptonic energy budget requires, at least, $10^{50}~\ergs$ of electrons within the SNR. Therefore a pure impulsive leptonic model of \HESSsource being the accelerator of high energy particles resulting in the GeV gamma-radiation as observed by \fermi-LAT is rejected.
\par 
A leptonic-hadronic impulsive scenario requires leptonic interactions to produce $1\%$ of the GeV gamma-rays as seen by \fermi-LAT to reduce the total SNR leptonic energy budget to $10^{48}~\ergs$. This leaves $99\%$ of gamma-rays to be the result of hadronic interactions from SNR with energy budget of $5-720\times 10^{50}~\ergs$. Therefore an impulsive scenario considering a combination of hadronic and leptonic interactions producing the observed GeV gamma-rays can be rejected.

\subsubsection{PWN (continuous)}

We will now examine the pulsar wind nebula as the source of high energy particles.
\par 
The spin down power of the pulsar powering PWN \HESSsource is of order $10^{36}~\ergspersecond$. The spin down power of the pulsar is not constant over time; \cite{2006A&A...460..365A} has suggested that the high gamma-ray luminosity may indicate that the spin-down power was far greater in the past.
\par 
From Table \ref{tab:SED_sum_1825_hadronic} a hadronic continuous scenario requires injection luminosities of $1.8\times 10^{39}~\ergspersecond$ and $9.4\times 10^{38}~\ergspersecond$ for ages $21~\kiloyear$ and $40~\kiloyear$ respectively. This far exceeds the spin down power of PSR\,J1826-1334, rejecting this scenario. Considering a leptonic continuous scenario for ages of $21~\kiloyear$ and $40~\kiloyear$, all three GeV regions require a total of $\approx 10^{37}~\ergspersecond$ in injection luminosity. If the spin down power of PSR\,J1826-1334 was greater in the past as suggested by \cite{2006A&A...460..365A}, GeV-ABC may be a reflection of an earlier epoch in the PWN history. The original spin-down power, $\dot{E}_0$ of the pulsar is linked to the present spin down power $E\qty(t)$ through:
\begin{align}
    \dot{E}\qty(t)&=\dot{E}_0\qty(1+\frac{t}{\tau_0})^{-\frac{n+1}{n-1}}
\end{align}
where $n$ is the braking index of the pulsar and $\tau_0$ is the initial spin-down timescale \citep{1973ApJ...186..249P}. The spin-down timescale can be determined from:
\begin{align}
    \tau_0 = \frac{P_0}{\qty(n-1)\abs{\dot{P_0}}}
\end{align}

Taking the assumption from \cite{2020A&A...640A..76P} that $\dot{P_0}=15~\si{\milli\second}$ and assuming $\dot{P}=\dot{P_0}$ with a braking index of $3$, the original spin-down power of PSR\,J1826-1334, $\dot{E_0}$, was in the order of $10^{39}~\ergspersecond$. This exceeds the injection luminosity for a leptonic scenario with the PWN as the accelerator of high energy particles. 
Electrons injected into the PWN by the pulsar are transported by a combination of advection and diffusion. At the edge of the PWN, it can be assumed that the electrons escape isotropically. Consequently the GeV gamma-ray emission towards GeV-ABC is expected to follow the photon fields through IC interactions. As the CMB photon field is constant, only the IR photon field would affect the morphology of gamma-ray emission. As seen in Fig.\,\ref{fig:IR_8_micron} the peaks in the GeV gamma-ray emission do not correspond to the IR field. Under this scenario, a preferential direction would be required for the advection/diffusion of electrons from the PWN.
\par 
\par 
Fig.\,\ref{fig:Particle_Transport} shows diffusive particle transport of electrons travelling a distance of $140~\pc$ in ambient density of $n=1~\centimeterminusthree$ versus the cooling of synchrotron and IC processes. This is equivalent to the distance that electrons travel after being emitted by the pulsar to reach GeV-B. the vertical lines represent the equivalent minimum and maximum electron energy seen by \fermi-LAT and \HESS respectively. Fast diffusion ($\chi = 1.0$) is required for electrons in this energy range to reach GeV-B within the age of PWN.
\par 
The High Altitude Water Cherenkov Observatory (HAWC) has observed $\gamma$-rays greater than $100~\TeV$ \citep{2019arXiv190908609H} suggesting that $E_{\text{e, max}}$ is greater than shown in figure \ref{fig:Particle_Transport}. The maximum electron electron able to reach GeV-B is determined by the intersection of diffusion time and the cooling time, i.e. where all electrons have lost their energy through leptonic interactions. On the other hand the minimum electron energy is represented by the intersection of diffusion time and the age of the pulsar. It can be concluded that for slow diffusion ($\chi=0.01$) no electrons are able to reach GeV-B; while for fast diffusion, electrons greater than $\approx 10~\TeV$ can travel to GeV-B in time. This is reaching the cutoff energy required to reproduce the spectral energy distribution of leptonic process as seen in Table\,\ref{tab:SED_sum_1825_leptonic}.
\par 
A more powerful pulsar can convert more of its spin down power into electron energy, allowing a greater proportion of higher energy electrons. This, in turn, allows more electrons to reach GeV-ABC in time to emit $\GeV$ radiation. Therefore unless advection or fast diffusion is considered or the PWN is powerful, electrons are unable to reach GeV-ABC from PSR\,J1826-1334 without significant energy losses.

\subsection{LS\,5039 as a particle accelerator} \label{sec:5039_discussion}

We will now discuss the possibility of LS\,5039 as the accelerator for high energy particles resulting in gamma-rays observed towards GeV-ABC.

\subsubsection{Progenitor SNR (Impulsive)}

From Table\,\ref{tab:SED_sum_5039_hadronic}, if GeV-ABC is the result of hadronic interactions from an impulsive progenitor SNR, no age of LS\,5039 can simultaneously explain the $\GeV$ emission as total energy budgets exceed $10^{50}~\ergs$. Due to the denser cloud towards GeV-B as seen in Fig.\,\ref{fig:CO_data} for all three clouds to be explained by the same source of high energy particles, the cosmic ray density must be approximately a factor of $10$ larger in GeV-B than GeV-A and GeV-C. It can be concluded that an impulsive hadronic source of cosmic rays from LS\,5039 cannot simultaneously explain the $\GeV$ regions observed by \cite{Araya:2019dez}. Similarly an impulsive leptonic source for any age of LS\,5039 cannot explain any of the $\GeV$ emission from GeV-ABC due to energy budgets exceeding $10^{49}~\ergs$ as shown in Table. \ref{tab:SED_sum_5039_leptonic}.

\subsubsection{Accretion powered (continuous)}

Microquasars such as LS\,5039 are considered to be candidates for particle acceleration up to gamma-ray energies \citep{2005Sci...309..746A}. The average accretion luminosity of LS\,5039 is \textit{\textbf{$8.0\times 10^{35}~\ergspersecond$ }}\citep{2005MNRAS.364..899C}. Approximately one third of accreted energy is radiated in the relativistic jet \citep{2005MNRAS.364..899C}. The remaining $5.3\times10^{35}~\ergs$ is assumed to be channelled into GeV-ABC through a jet. It is unknown whether this jet is hadronic or leptonic in origin. This jet is a continuous source of particles into the region towards GeV-ABC. While the jet generally points in the direction of GeV-B (see Fig\,\ref{fig:H_alpha}), the precession of the jet may allow electrons to be channelled into GeV-A and GeV-C. Particles may also diffuse from the jet escaping into the necessary regions.
\par 
A hadronic scenario requires a total injection luminosity into GeV-ABC of $10^{39}-10^{36}~\ergspersecond$ for an age range of $10^{3}-10^{6}~\si{yrs}$. All ages require total injection luminosity greater than the accretion luminosity can provide; rejecting a hadronic accretion powered scenario.
\par 
On the other hand a leptonic scenario requires a total injection luminosity into GeV-ABC of $10^{38}-10^{36}~\ergspersecond$ for an age range of $10^{3}-10^{6}~\si{yrs}$. The younger ages of $10^3~\si{yrs}$ and $10^5~\si{yrs}$ can be rejected. All scenarios can vary systematically up to $56\%$ in injection luminosity, an age of $10^6~\si{yrs}$ can provide the energetics required to reproduce the gamma-rays as seen by \fermi-LAT. But this age is greater than the age of $\approx10^5~\si{yrs}$ predicted by \cite{2012A&A...543A..26M}. Therefore a leptonic scenario with a continuous jet powered by the accretion onto compact object in LS\,5039 can be rejected.
\par 
Using the calculated hydrogen densities towards the regions of interest in LS\,5039 in the $40-60~\kmpersec$ range rather than the $15-30~\kmpersec$ range will not alter the results due to values being within a factor of 10 of each other.
\par 
In summary, it is unlikely that LS\,5039 is the source of the new region of GeV gamma-ray emission. 

\subsection{Combination of LS\,5039 and \HESSsource}

The new region of GeV gamma-rays may be a line of sight combination of \HESSsource and LS\,5039. As discussed in section \ref{sec:1825_discussion} and \ref{sec:5039_discussion} a hadronic scenario requires cosmic ray energy density to be ten times higher in GeV-A and GeV-C compared to GeV-B assuming the dense gas observed in the velocity range 15-30 km/s in Fig.\,\ref{fig:CO_data} lies at the same distance as HESS J1825-137. Note in the case of HESS\,J1825-137, it assumes the dense gas observed in $15-30~\kmpersec$ range in Fig.\,\ref{fig:CO_data} lies at the same distance as HESS\,J1825-137. If the GeV gamma-ray emission from GeV-A and GeV-C is unrelated to emission from GeV-B, this issue will be negated. 
\par 
As seen in Fig.\,\ref{fig:H_alpha} the region around GeV-ABC contains several astrophysical environments; a H$\alpha$ region believed to be associated with the SNR linked to \HESSsource and a relativistic jet from LS\,5039. Even though \HESSsource and LS\,5039 are at different distances ($3.9~\kpc$ and $2.5~\kpc$ respectively), the combination of these two processes may explain the spectra observed by \fermi-LAT.
\par 
Peaks GeV-B and GeV-C have similar spectral indices, $\Gamma=1.7\pm 0.4$ and $\Gamma=1.78\pm 0.25$ respectively, indicating a common origin scenario, whilst GeV-A has a marginally harder spectra with $\Gamma = 1.43\pm0.23$. GeV-A is positioned the closest to both \HESSsource and LS\,5039. As shown by equation \ref{eq:R_diffusion} and \ref{eq:Diffusion}, high energy particles are able to travel further distances than lower energy particles in the same time. Clouds closer to the source of high energy particles are expected to have a softer spectrum than clouds lying further from the source for both continuous and impulsive sources \citep{1996A&A...309..917A}. This is the opposite to what is observed in GeV-ABC.

\subsection{Particle Accelerators unrelated to \HESSsource and LS\,5039} \label{sec:other_accelerators}

\begin{figure}
    \centering
    \includegraphics[width=\columnwidth]{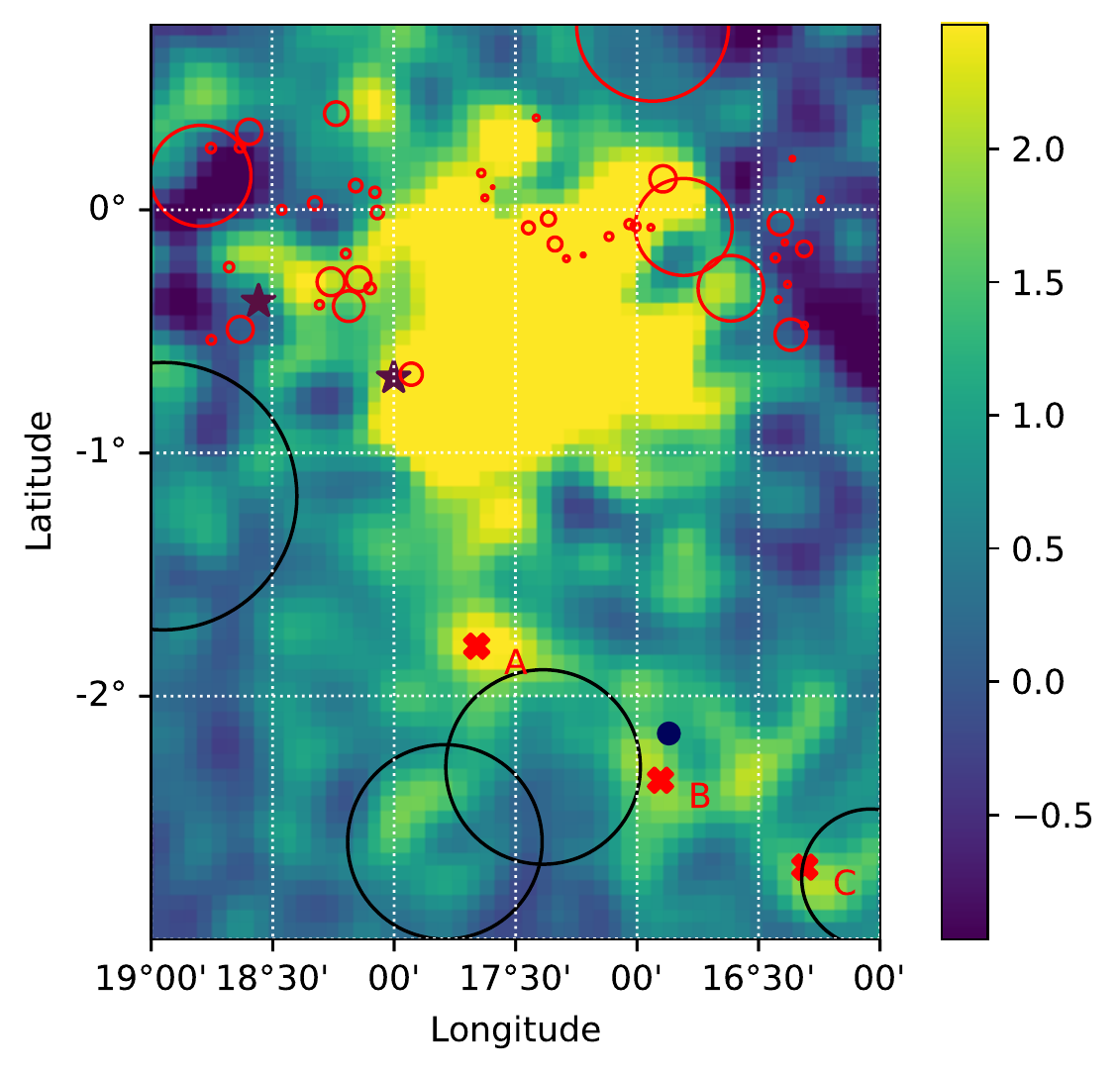}
    \caption{\fermi-LAT count map above $10~\GeV$ \citep{Araya:2019dez} towards \HESSsource is shown with alternative accelerators for high energy particles towards HESS\,J1825-137 and GeV-ABC. Red circles describe HII regions where star formation may occur as given by the WISE catalogue \citep{2014AAS...22331201A}. Black circles show the location of other SNR towards the region of interest as described in section \ref{sec:other_accelerators}. Dark purple stars represent pulsars PSR\,J1826-1334 and PSR\,J1826-1256 \citep{2005AJ....129.1993M}. The water maser, G016.8689-02.1552, can be seen as a dark blue dot nearby GeV-B \citep{2011MNRAS.418.1689U}}
    \label{fig:other_accelorators}
\end{figure}

\begin{figure}
    \centering
    \includegraphics[width=\columnwidth]{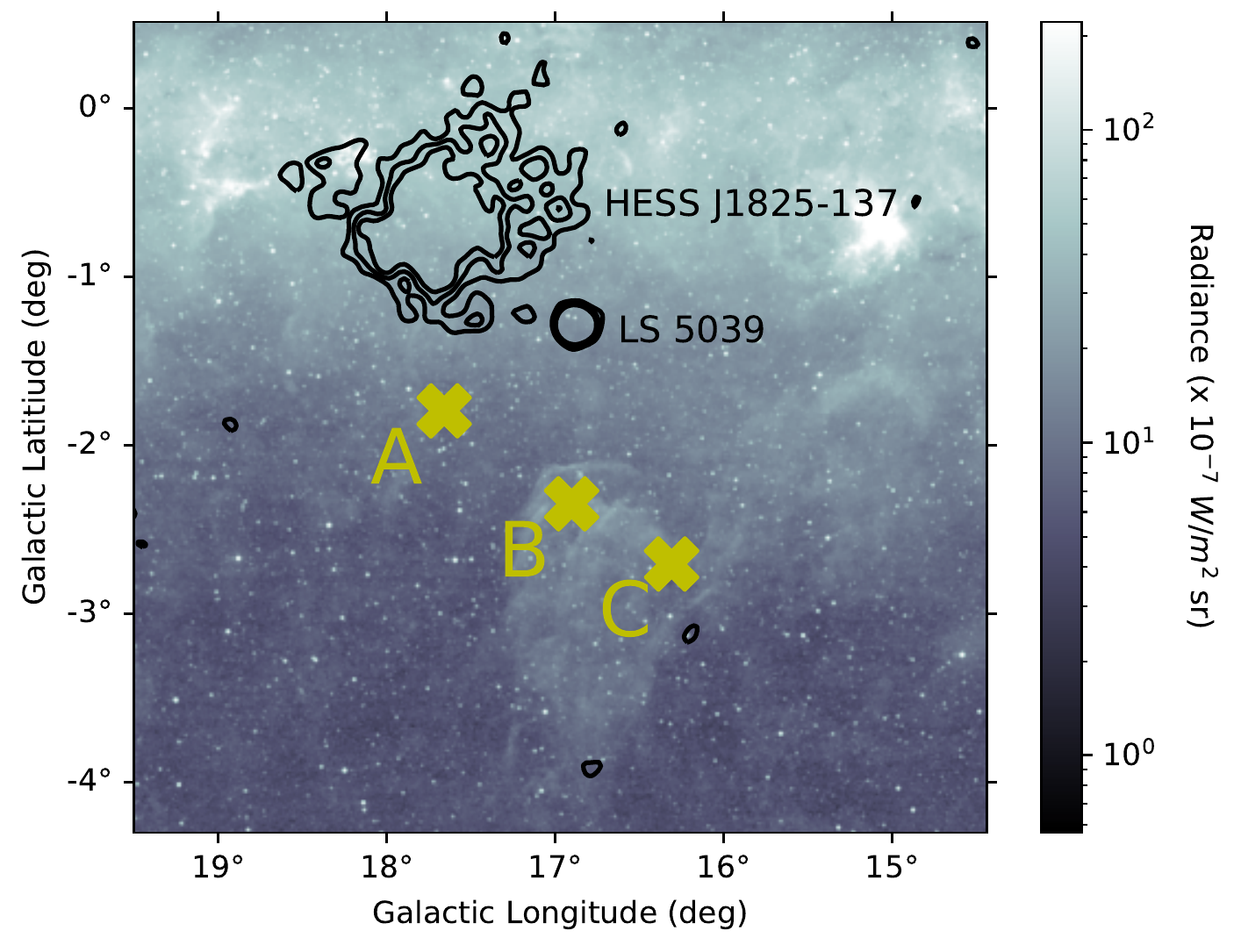}
    \caption{Infrared emission in the $8.26~\si{\micro\meter}$ band towards GeV-ABC \citep{2003AAS...203.5708E}. Overlayed are the HESS significance contours towards \HESSsource at $1\sigma$, $2\sigma$ and $3\sigma$ \citep{2018A&A...612A...1H}.}
    \label{fig:IR_8_micron}
\end{figure}

Towards GeV-ABC, there are four known supernova remnants; these are SNR G017.4-02.3, SNR G018.9-01.1, SNR G016.2-02.7 and SNR G017.8-02.6; see figure \ref{fig:other_accelorators}. From equation \ref{eq:cooling_time_sync}, the cooling time of electrons resulting in synchrotron emission is proportional to the energy; as the supernova remnant ages, higher energy electrons escape from the system or lose their energy decreasing the emitted photon energy. Therefore as a supernova remnant ages, the amount of X-ray emission detected decreases. Three of the four supernova remnants have no current X-ray detection, indicating that these SNRs are old (at least in the later stages of the Sedov-Taylor phase). They are therefore unlikely to be a source of high energy particle acceleration, resulting in the production of $\GeV$ gamma-rays. The remaining supernova remnant, SNR G18.9-1.1, has a partial X-ray shell \citep{2004ApJ...603..152H}. Based on radio measurements by \cite{2004ApJ...603..152H} it is located $2$ or $15.1~\kpc$ away. More recent research indicates a distance of $2.1\pm 0.4~\kpc$ and age of $3700~\si{yr}$ \citep{2019arXiv191005407R}. As mentioned by \cite{Araya:2019dez}, if GeV-ABC is the result of a combination of SNRs, \fermi-LAT images will show distinct sources above $10~\GeV$ with the given \fermi-LAT resolution.
Star forming regions have also been suggested as an accelerator of cosmic rays. See Fig.\,\ref{fig:other_accelorators} to see location of star forming regions, SNRs and pulsars towards HESS\,J1825-137 and GeV-ABC. The presence of water maser G016.8689-02.1552, as shown in Fig. \ref{fig:other_accelorators}, highly suggests star formation towards this region \citep{2011MNRAS.418.1689U}. This is supported by data from the MSX satellite; data reveals infra-red emission towards GeV-B and GeV-C in the $8.26~\si{\micro\meter}$ band (see Fig. \ref{fig:IR_8_micron}).

\section{Conclusions}
This study presented spectral models of a region of GeV gamma-ray emission to the south of HESS J1825-137 revealed by \fermi-LAT. Different accelerators were proposed to be an origin for high energy particles that created this new region of gamma-rays; the PWN (continuous) and SNR (impulsive) associated with HESS\,J1825-137, and the binary system and microquasar LS\,5039 (continuous) as well as the associated progenitor SNR (impulsive). We found that the progenitor SNR related to HESS\,J1825-137 is unlikely to be the sole source of high energy particles due to the energetics needed to replicate the SED is greater than what the system can provide. For example an impulsive SNR releases approximately $10^{50}~\ergs$ of cosmic rays (with $1\%$ of energy channeled into electrons), whereas the SED model of the progenitor SNR of \HESSsource is required to provide either $10^{52}~\ergs$ of protons or $10^{50}~\ergs$ of electrons to replicate the spectral energy distribution. A continuous acceleration scenario from the PWN (powered by the pulsar) into GeV-ABC requires particle injection luminosity to be of order $~10^{39}~\ergspersecond$ and $~10^{37}~\ergspersecond$ for hadronic and leptonic particles respectively. GeV-ABC may be a reflection of an earlier epoch in the PWN history, where the pulsar was more powerful. Therefore the PWN may be a possible accelerator for high energy electrons resulting in this new region of gamma-ray emission, assuming fast diffusion perhaps including advection towards this region. 
Moreover, it is unlikely that leptonic inverse-Compton emission into this region will produce the localised features such as GeV-ABC.
LS\,5039 at any age cannot solely explain the GeV emission from GeV-ABC with required injection luminosity > $10^{36}~\ergspersecond$ compared to the $10^{35}~\ergspersecond$ accretion luminosity of LS\,5039 \citep{2005MNRAS.364..899C}. However a combination of emission from both \HESSsource and LS\,5039 could be the cause of the gamma rays.

\section*{Acknowledgements}

This research has made use of the NASA's Astrophysics Data System and the SIMBAD database, operated at CDS, Strasbourg, France. T.C. acknowledges support through the provision of Australian Government Research Training Program Scholarship.

\section*{Data Availability}

No new data were generated or analysed in support of this research.




\bibliographystyle{mnras}
\bibliography{Main_Document} 



\appendix

\section{ISM parameters}

\floatstyle{plaintop}
\restylefloat{table}

\begin{table}[H]
    \caption{Calculated molecular parameters for \HESSsource and the new GeV emission regions GeV-A, GeV-B, GeV-C as shown in Fig.\,\ref{fig:CO_data}. $M_H$ and $n_H$ describes the average mass and density respectively over the new GeV regions.}
	\centering
	\begin{tabular}{cccc}
		\hline
		$15-30~\kmpersec$ & Region & $M_{H}$ ($M_\odot$) & $n_H$ ($\centimeterminusthree$) \\
		\hline
		& \HESSsource & $1.14\times10^5$ & $39$ \\
		& GeV-A & $3.67\times 10^3$ & $5$\\
		& GeV-B & $1.36\times10^5$ & $79$ \\
		& GeV-C & $1.53\times 10^3$ & $2$ \\
		\hline
		$40-60~\kmpersec$ & Region & $M_{H}$ ($M_\odot$) & $n_H$ ($\centimeterminusthree$) \\
		\hline
		& \HESSsource & $5.18\times 10^5$ & $176$ \\
		& GeV-A & $8.49\times 10^3$ & $13$ \\
		& GeV-B & $1.21\times 10^4$ & $7$ \\
		& GeV-C & \multicolumn{2}{c}{\textit{No ISM values}} \\
		\hline
	\end{tabular}
	\label{tab:CO_density}
\end{table}

\begin{table}[H]
    \caption{Calculated HI densities for \HESSsource and the new GeV emission regions GeV-A, GeV-B and GeV-C.}
	\centering
	\begin{tabular}{cccc}
		\hline
		$15-30~\kmpersec$ & Object & $M_{H}$ ($M_\odot$) & $n_H$ ($\centimeterminusthree$) \\
		\hline
		& \HESSsource & $3.53\times10^3$ & $1.2$ \\
		& GeV-A & $8.91\times 10^2$ & $1.4$ \\
		& GeV-B & $1.51\times 10^3$ & $0.9$ \\
		& GeV-C & $7.70\times 10^2$ & $1.2$ \\
		\hline
		$40-60~\kmpersec$ & Object & $M_{H}$ ($M_\odot$) & $n_H$ ($\centimeterminusthree$) \\
		\hline
		& \HESSsource & $4.29\times 10^3$ & $1.5$ \\
		& GeV-A & $4.36\times 10^2$ & $0.7$ \\
		& GeV-B & $5.06\times 10^2$ & $0.3$ \\
		& GeV-C & $3.11\times 10^2$ & $0.5$ \\
		\hline
	\end{tabular}
	\label{tab:HI_density}
\end{table}

\begin{table}[H]
    \caption{Calculated H$\alpha$ densities for \HESSsource and new GeV emission regions.}
	\centering
	\begin{tabular}{ccc}
		\hline
		 Object & Method A ($\centimeterminusthree$) & Method B ($\centimeterminusthree$) \\
		\hline
		\HESSsource & $8.88\times 10^{-6}$ & $4.12\times 10^{-6}$ \\
		GeV-A & $1.12\times 10^{-6}$ & $2.40\times 10^{-6}$ \\
		GeV-B & $6.49\times 10^{-6}$ & $5.13\times 10^{-6}$ \\
		GeV-C & $2.45\times 10^{-6}$ & $5.23\times 10^{-6}$ \\
		\hline
	\end{tabular}
	\label{tab:Halpha_density}
\end{table}

\section{\halpha density calculation method} \label{sec:Halpha_method}

\subsection*{Method A}

Method A assumes that the density of photons in the region of interest is approximately equal to the density of ionised gas $n\approx u_\text{ph}$. This assumes that atoms are not being re-excited by an external source. Considering a spherical shell located at distance $d$ from the source with thickness $\dd{\ell}$; the volume of the shell is given by $\dd{V}=4\pi d^2 \times \dd{\ell}$. Photons emitted by the source travel at the speed of light, therefore $\dd{\ell}=c\dd{t}$. The number of photons emitted by the source in time $\dd{t}$ is related to the luminosity $L$ through $\dd{N}=L\dd{t}$. Using the original approximation, the density of ionised hydrogen in a region of interest:

\begin{align}
    n\approx u_\text{ph}=\dv{N}{V}=\frac{L}{4\pi d^2 c}
\end{align}Let the region of interest have solid angle $\Omega$ and lying at distance $d$. The luminosity of the region is given by:

\begin{align}
    L~[\si{photon\per\second}] = \frac{d^2}{10^{-10}} \Omega I
\end{align}

where $I$ is the measured \halpha intensity in Rayleigh units. 

\subsection*{Method B}

Method B considers basic radiation transfer. The density of atoms in the $i$th excited state emit photons at frequency $\nu$ through spontaneous emission is related to the emission coefficient by:

\begin{align}
    n_i&=\frac{j_\nu \Omega_\text{Earth}}{E_\nu A\phi (\nu)} \label{eq:h_alpha_dens_2}
\end{align}
where A is the Einstein coefficient, $\Omega_\text{earth}$ is the solid angle of Earth projected at source lying at distance $d$ and $\phi(\nu)$ is the spectral line shape normalised by:
\begin{align}
    \int \phi(\nu)=1
\end{align}
Assuming that hydrogen atoms in the $n=3$ state emit mainly H$\alpha$ light; $\phi=0$ in all frequencies except when $\nu=\nu_{H\alpha}$. The photon radiance $L_\text{rad}$ is related to the intensity $I$ in Rayleigh's through:

\begin{align}
    L_\text{rad}~[\si{photons\per\meter\squared\per\second\per\steradian}] &= \frac{L}{4\pi d^2}
\end{align}
The photon intensity can be found utilising $I_\nu=L\frac{E_\nu}{\nu}=hL$ where h is Planck's constant. Let $s$ be the thickness of gas in the line of sight and assuming the emission coefficient is constant, the emission coefficient and intensity are related by:
\begin{align}
    j_\nu = \frac{I_\nu}{s}
\end{align}
This can be used in combination with equation \ref{eq:h_alpha_dens_2} to obtain the photon density.
\par 

\section{SED Model} \label{sec:newsedprod}

The SED modelling code includes various astrophysical processes; included proton-proton interactions:

\begin{align}
    &\proton + \proton \rightarrow \pionplus + \pionminus + \pionneutral \\
    &\pionneutral \xrightarrow{decay} \gamma + \gamma
\end{align}

Inverse Compton interactions:

\begin{align}
    \ce{e^{-*}} + \gamma^* \rightarrow \electron + \gamma
\end{align}

Bremsstrahlung interactions with a nucleus with proton number $Z$:

\begin{align}
    \ce{e^{-*}} + Z \rightarrow \gamma + \electron + Z
\end{align}

and synchrotron interactions:

\begin{align}
    \ce{e^{-*}} + \Vec{B} \rightarrow \electron
\end{align}

The evolution of the cosmic ray energy distribution with Lorentz factor $\gamma$ at time $t$ is given by:
	
	\begin{align}
		\pdv{n\qty(\gamma,t)}{t}&=\pdv{ }{\gamma}\qty[\dot{\gamma}\qty(\gamma)n\qty(\gamma,t)]+S\qty(\gamma,t) \label{eq:evol}
	\end{align}
	
	where $S\qty(\gamma,t)$ is the source term, $\dot{\gamma}\qty(\gamma)$ represents the energy loss rate of a particle with Lorentz factor $\gamma$. The analytical solution of equation \ref{eq:evol} is:
	
	\begin{align}
		n\qty(\gamma, t)&=\frac{1}{\dot{\gamma}}\int_\gamma^{\gamma_0} S\qty(\gamma^{''},t-\tau\qty(\gamma^{''}))\dd{\gamma^{''}}+\frac{\dot{\gamma}_0}{\gamma} n\qty(\gamma_0,0) \label{eq:model}
	\end{align}
	
	where $\tau$ is a variable describing the time for a cosmic ray with initial Lorentz factor $\gamma^{'}$ to evolve to factor $\gamma$:
	
	\begin{align}
		\tau(\gamma^{'},\gamma)=\int_\gamma^{\gamma^{'}}\frac{\dd{\gamma^{''}}}{\dot{\gamma}\qty(\gamma^{''})}
	\end{align}
	
	and $\gamma_0$ is the initial Lorentz factor. The code solves equation \ref{eq:model} considering hadronic and leptonic interactions and then extracts the spectral energy distribution. The model allows the user to choose whether the case is leptonic, hadronic or a mixture. Similarly the user can choose if the model is continuous (constant cosmic ray input, eg a PWN) or impulsive (releases all the cosmic rays at once, eg a SNR). Other parameters such as the age and distance from the source, density and magnetic field of background material, the total energy and spectral distribution of cosmic rays and background photon field energy distribution can be changed depending on the source.
	
	~
	
	To find the SED at time $t$, for each lorentz factor $\gamma$ the lorentz factor at earlier time, $\gamma_0$ is derived. In the case of an impulsive source $\gamma_0$ is simply $\gamma$ at $t=0$. The total cooling rate is given by \cite{2007A&A...474..689M}:
	
	\begin{equation}
	    \begin{aligned}
		 \dot{\gamma}\qty(\gamma) = .
		    \begin{cases}
		         b_s\gamma^2+b_c\qty(3\ln\gamma+18.8)+5.3b_b + \\
		         \quad\sum\limits_{t=i} b_{\text{IC}}^i\gamma^2 F_{\text{KN}}^i\qty(\gamma),     &\qquad \text{for leptonic cases}\\
		        \frac{1}{n_H c\sigma_{\text{pp}\qty(\gamma)}}, &\qquad           \text{for hadronic cases}
	    	\end{cases}
	     \end{aligned}
	\end{equation}
	
	where:
	
	\begin{itemize}
		\setlength\itemsep{1em}
		\item $b_s=1.292\times 10^{-15}\qty(B/10^3\si{\micro\gauss})^2~\si{\per\second}$ is a synchrotron loss constant.
		\item $b_c=1.491\times 10^{-14} \qty(n_H/1\centimeterminusthree)$ is the Coulomb loss constant.
		\item $b_b=1.37\times 10^{-16}\qty(n_H/1\centimeterminusthree)~\si{\per\second}$ is the Bremsstrahlung loss constant.
		\item $b_\text{IC}=5.204\times 10^{-20}\qty(u_0^i/\si{\electronvolt})~\si{\per\second}$ is a IC loss constant with the energy density of photons given by $u_0$.
	    \item $\sum_{t=i}$ sums over all radiation fields contributing to the Inverse compton gamma-ray flux.
		\item $n_H$ is the density of the ambient hydrogen gas.
		\item $\sigma_{\text{pp}\qty(\gamma)}$ is the cross section for proton-proton interactions.
	\end{itemize}
	
	~
	
	To obtain $\gamma_0$, the following two steps are repeated until $t=t_\text{age}$:
	
	\begin{enumerate}
		\item derive $\Delta t=\dd{\gamma}/\dot{\gamma}\qty(\gamma)$
		\item Increment $\gamma$ by $\dd{\gamma}$
	\end{enumerate}

	with automatic adjustion of the $\dd{\gamma}$ step. Another parameter in the code is ``escape". If this parameter is activated, once a particle escapes the system, it is no longer considered. The final synchrotron flux is given by:
	
	\begin{align}
		P\qty(\nu)&=\frac{\sqrt{3}e^3B}{mc^2} \frac{\nu}{\nu_c}\int_{\frac{\nu}{\nu_c}}^\infty K_\frac{5}{3}(x) \dd{x}
	\end{align}
	
	where $e$ and $m$ are the charge and mass of an electron respectively, $\nu$ is the frequency of the gamma-ray, $\nu_c$ is the critical frequency of the emission, and $K_\frac{5}{3}$ is the modified Bessel Function. The final Inverse Compton flux radiated by a single electron with energy $\epsilon$ is given by:
	
	\begin{align}
		\dv{N}{E_\gamma}&=\frac{3}{4} \sigma_T c\int \frac{n\qty(\epsilon)\dd{\epsilon}}{\epsilon} F_{\text{KN}}\qty(E_e, E_\gamma, \epsilon)
	\end{align}
	
	where $\sigma_T$ is the Thompson cross section and $F_{\text{KN}}$ is the Klein-Nishina cross section. The final Bremsstrahlung Flux is given by:
	
	\begin{align}
		\dv{N}{E_\gamma}= nc \int \dd{\sigma}\qty(E_e, E_\gamma, Z)\dd{E_e}
	\end{align}
	
	where $Z$ is the atomic number of the target material and $\dd{\sigma}$ is defined in \cite{1970RvMP...42..237B}.  Finally proton-proton interactions produce a flux of:
	
	\begin{align}
		\dv{N}{E_\gamma}= nc\int_{E_p=E_\gamma}^\infty A_\text{max}\qty(T_p)F(E_\gamma, T_p)\dd{E_p} \label{eq:app_proton_proton_flux}
	\end{align} 
	
	where $n$ is the density of protons, $A_\text{max}(T_p)$ is the pion production cross section, $T_p$ is the kinetic energy of the proton and $F(E_\gamma,T_p)$ is the spectra of gamma rays emitted for a single proton of energy $E_p$.

\section{Other Spectral Energy Distribution Plots}

\begin{figure*}
	\begin{subfigure}[b]{\columnwidth}
        \includegraphics[width=\textwidth]{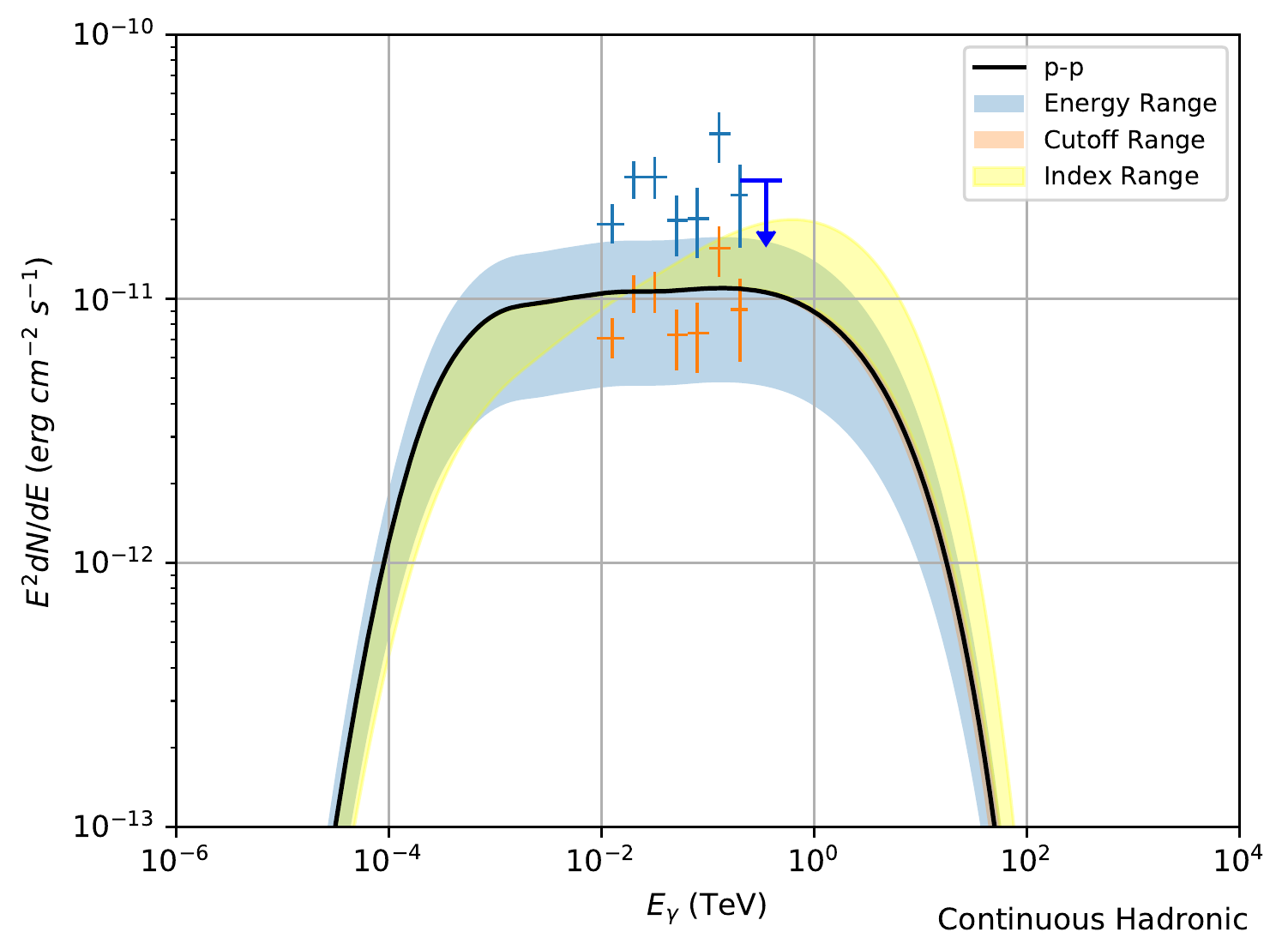}
	\end{subfigure}
	 ~
	\begin{subfigure}[b]{\columnwidth}
	    \includegraphics[width=\textwidth]{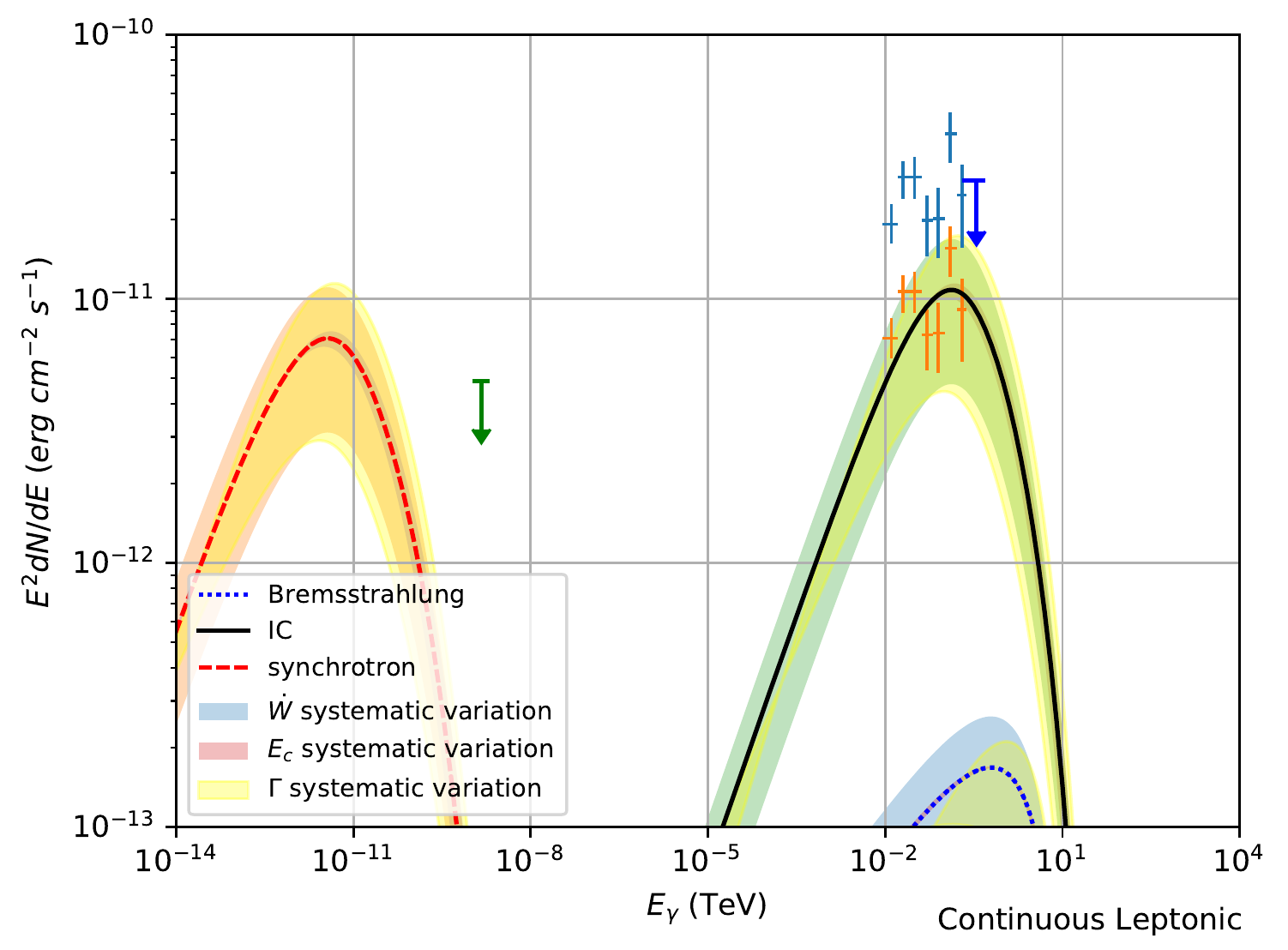}
	\end{subfigure}
    \caption{SED example for GeV-A with continuous hadronic SED (\textit{left}) and continuous leptonic SED (\textit{right}). This assumes that LS\,5039 (at age $10^5~\si{yrs}$) is the source of acceleration. The upper data points represent the total SED as measured by \fermi-Lat. The lower data points is $36.9\%$ of this spectra due to source being only GeV-A. The green cross is the ROSAT X-ray upper limit towards GeV-ABC. The blue, pink and yellow shaded regions represent the systematic variation of energy budget ($W$) or injection luminosity ($\dot{W}$), cutoff energy and spectral index respectively. See Tables\,\ref{tab:SED_sum_5039_hadronic} and \ref{tab:SED_sum_5039_leptonic} for input parameters.}
    \label{fig:example_SED_ls5039}
\end{figure*}


\bsp	
\label{lastpage}
\end{document}